\documentclass{ws-ijmpa}
\usepackage[super,compress]{cite}
\usepackage{graphicx}
\usepackage{amsmath}
\usepackage{tabularx}
\usepackage{enumitem}
\usepackage{epsf}
\usepackage{float}
\usepackage{array,booktabs}
\usepackage{epstopdf}
\usepackage{footnote}
\usepackage{threeparttable}
\usepackage[paperwidth=210mm,paperheight=297mm,centering,hmargin=2.5cm,vmargin=3cm]{geometry}
\usepackage{etoolbox}
\AtBeginEnvironment{bmatrix}{\setlength{\arraycolsep}{5pt}}

\newcolumntype{P}[1]{>{\centering\arraybackslash}p{#1}}
\allowdisplaybreaks[1]
\newcommand\numberthis{\addtocounter{equation}{1}\tag{\theequation}}

\begin{document}

\markboth{S. M. Zebarjad and S. Zarepour}{Two-body decay widths}

%
%

\title{%
\hfill{\normalsize\vbox{%
\hbox{\rm }
 }}\\
{Two-body decay widths of lowest lying and next-to-lowest lying scalar and pseudoscalar mesons in generalized linear sigma model}}

\author{S. Mohammad Zebarjad\footnotemark[1] \, and Soodeh Zarepour\footnotemark[2]  }

\address{Department of Physics,
 Shiraz University,\\
  Shiraz 71454, Iran\\
\footnotemark[1]  zebarjad@susc.ac.ir\\
\footnotemark[2]  soodehzarepour@shirazu.ac.ir}

\maketitle


\begin{abstract}
Two-body decay widths of lowest lying and next-to-lowest lying scalar and pseudoscalar mesons are studied
in  Generalized Linear Sigma Model (GLSM) of low-energy QCD. This model which considers mixing between
``two quark" and ``four quark" chiral nonets
 has been employed to investigate various decays and scatterings in low energy region of QCD.
In this paper, $\Gamma[f_0(980) \rightarrow K \bar{K}]$ and $\Gamma[a_0(980) \rightarrow K \bar{K}]$ are obtained and it is shown that two-body decay widths of lowest lying mesons are well predicted by this model while for the next-to-lowest lying mesons, only some of the decay widths agree with the experimental results.
 We have  compared the predicted decay widths in GLSM with the results obtained in single nonet linear sigma model (SNLSM) to indicate that chiral nonet mixing  greatly improves the predictions of SNLSM  for decay widths.

\end{abstract}

\ccode{PACS numbers: 13.25.-k, 12.39.Fe}

\maketitle
\section{Introduction}
\noindent
Since low energy QCD is nonperturbative, the usual field theory methods are impractical and researchers
have been examining effective field theory approaches. In theses approaches like chiral perturbation theory \cite{ch1,ch2,ch3,ch4,ch5,ch6,ch7},
 linear and non-linear sigma models \cite{lnsm1,lnsm2,lnsm3,lnsm4,lnsm5,lnsm6}, extended linear sigma model \cite{elsm1,elsm2,elsm3} and generalized linear sigma model (GLSM)
 \cite{g1,g2,g4,global}, the underlying chiral symmetry of QCD is used in order to construct hadronic Lagrangians.
\par
Generalized linear sigma model is successful in describing some unusual features of scalar and pseudoscalar mesons below
2 GeV by allowing mixing between a quark-antiquark nonet and a four-quark nonet. As we know, the light scalar nonet
has unusual spectroscopy compared to the vector one. The vector mass ordering corresponds to the number of $s$-type
quarks in each state, i.e.,  $m[\rho(776) \sim u\bar{d}]< m[K^*(892) \sim u\bar{s}]<m[\phi(1020) \sim s\bar{s}]$, while this is not the case for the scalars. In fact, the conventional $q\bar{q}$, $SU(3)$ picture doesn't
match with the mass of the scalar mesons below 1 GeV. For example, the isodoublet $K_0^*(800)$ should be heavier than the isovector $a_0(980)$ but: $m[K_0^*(800)\sim u\bar{s}]<m[a_0(980)\sim u\bar{d}]$. In 1977, Jaffe
 examined the possibility that the light scalar mesons are made of two quarks and two anti-quarks (MIT bag model) and showed that this proposal can explain the
unusual mass spectrum of the scalar mesons \cite{jaffe1,jaffe2,jaffe3}. In this picture the mass ordering of the scalar mesons becomes proportional to the number of $s$-type
quarks in each state. As an example, $K_0^*$ with one strange quark becomes lighter than $a_0(980)$ with two strange quarks,  $m[K_0^*(800)\sim \bar{d}\bar{s}ud]<m[a_0(980)\sim \bar{d}\bar{s}su]$, which is consistent with the experimental data. Others also studied properties of light scalar mesons  in tetraquark picture (see Refs. \citen{Maiani,Pelaez1,Pelaez2}).
 \par
 Since this pure four-quark picture is deficient in describing some decay properties of lowest lying scalars and besides the fact that the pure quark-antiquark picture for the next-to-lowest lying scalars can not completely describe their properties, mixing among these states was considered as a solution to this puzzling feature \cite{mix1,mix2,mix3,mix4,mix5,mix6,mix7,mix8,mix9,mix10}. Mechanism for such mixing in a nonlinear chiral Lagrangian was studied in Ref. \citen{me} and it was shown that due to level repulsion, the lowest lying scalar mesons become lighter and this gives an answer to the question of why scalar mesons are lighter than the axial vector mesons.
 \par
 The effect of possible mixing among two quark and four quark chiral nonets on the masses and some decay properties of scalar and pseudoscalar mesons, was studied in generalized linear sigma model and it was found that the scalar mesons below 1 GeV are predominantly four-quark states  while the lowest lying pseudoscalar mesons are closer to $q\bar{q}$ type\cite{global}. The role of these mixing patterns were also explored in $\pi \pi$ \cite{pipi}, $\pi K$ \cite{piK} and $\pi \eta$ \cite{pieta} scatterings and it was shown that the model prediction for the scattering amplitudes are reasonable and in good agreement with the experiment for the energy region below $1$ GeV. Also the poles in the scattering unitarized amplitudes, which correspond to the physical resonances, were found to represent the scalars with masses and decay widths close to experimental data. Furthermore, the $\eta'\rightarrow \eta \pi \pi$ decay was probed within this model and it was shown that the predicted decay width agrees with the experiment up to about $1\%$ \cite{etap}.
 \par
  In this article we will investigate two-body decays of lowest lying and next-to-lowest lying scalar and pseudoscalar mesons. For more convenience the  experimental data for masses and decay widths are given in Table \ref{table1} \cite{pdg}. A brief review of the model is given in Sec. \ref{sec2} and in Sec. \ref{sec3} the predictions of the model for all hadronic two-body decay widths of scalars and pseudoscalars below $2$ GeV are presented. Sec. \ref{sec4} is devoted to the unitarity corrections and their effects on decay widths. In Sec. \ref{sec5},  we summarize and discuss the results.

\begin{table}[!htbp]
\footnotesize
\centering
\label{exp.}
\tbl{Brief review of lowest lying and next-to-lowest lying scalar and pseudoscalar mesons properties \cite{pdg}.}
{\begin{tabular}{llllll}
\noalign{\hrule height 1pt}
\noalign{\hrule height 1pt}
particle& $I^G(J^{PC})$& mass (MeV)  & full width (MeV)& hadronic two-body & branching ratios or fractions $\Gamma_i/\Gamma_j $ \\
& & & &  decay modes& \\
\noalign{\hrule height 1pt}
\multicolumn{5}{l}{\textbf{\begin{small}\underline{below $1$ GeV}\end{small}}}\\
$\pi^0$&$1^-(0^{-+})$& $134.9766 \pm 0.0006$&  $\sim 10^{-6}$ &  ------ & ------\\
$\pi^{\pm}$&$1^-(0^-)$& $139.57018 \pm 0.00035$& $\sim 10^{-14}$ &  ------ & ------\\
$K^{\pm}$&$\frac{1}{2}(0^-)$& $493.677 \pm 0.016$& $\sim 10^{-14}$  &  $\pi^+\pi^0$& $(20.66 \pm 0.08)\%$\\
$K^0$&$\frac{1}{2}(0^-)$& $497.614 \pm 0.024$& ------  &------&------  \\
$f_0(500)$ or $\sigma$ &$ 0^+(0^{++})$& $400$ to $550$  & $400$ to $700$ &  $\pi \pi$ (dominant) & ------   \\
$\eta$ &$0^+(0^{-+})$& $547.862 \pm 0.018$ & $(1.31 \pm 0.05)\times 10^{-3}$& ------& ------ \\
$K_0^*(800)$ or $\kappa$&$\frac{1}{2}(0^+)$& $682\pm 29$& $547 \pm 24$ &  $\pi K$ & ------  \\
$\eta'$&$0^+(0^{-+})$& $957.78 \pm 0.06$ & $0.198 \pm 0.009$ & ------ & ------\\
$a_0(980)$&$1^-(0^{++})$ &  $980\pm 20$& $50 $ to $100$ &    $ \left\{\begin{tabular}{@{\ }l@{}}
 $\pi \eta$ (dominant)   \\  $K \overline{K}$   \end{tabular}\right.$ &$ \Gamma(K\overline{K})/\Gamma(\pi \eta)=0.183\pm 0.024 $ \\
 $f_0(980)$&$0^+(0^{++})$  & $990 \pm 20$  & $40$ to $100$ &     $ \left\{\begin{tabular}{@{\ }l@{}}
  $\pi \pi$ (dominant)  \\  $K \overline{K}$ \end{tabular}\right.$ & ------    \\
 \multicolumn{5}{l}{\textbf{\begin{small}\underline{above $1$ GeV}\end{small}}}\\
   $\eta(1295)$ &$ 0^+(0^{-+})$  & $1294 \pm 4$  & $55 \pm 5$ &     $ \left\{\begin{tabular}{@{\ }l@{}}
 $a_0(980) \pi$  \\  $\sigma \eta$ \end{tabular}\right.$  &  $ \begin{tabular}{@{\ }l@{}}
 $\Gamma(a_0(980)\pi) /\Gamma(\sigma \eta)=0.48 \pm 0.22$\end{tabular}$\\
 $\pi(1300)$&$1^-(0^{-+})$ & $1300 \pm 100 $& $200 \,{\rm to}\, 600$& $\rho \pi$&------ \\
 $f_0(1370)$&$0^+(0^{++})$   & $1200$ to $ 1500$  & $200$ to $500$ &     $ \left\{\begin{tabular}{@{\ }l@{}}
 $\pi \pi$ \\  $\pi(1300) \pi$ \\  $\eta \eta$ \\  $K \overline{K}$ \end{tabular}\right.$  & ------   \\
 $\eta(1405)$&$0^+(0^{-+})$   & $1408.9 \pm 2.4$  & $51.1 \pm 3.2$ &     $ \left\{\begin{tabular}{@{\ }l@{}}
 $a_0(980) \pi$  \\  $f_0(980) \eta$ \end{tabular}\right.$ & ------  \\
  $K_0^*(1430)$&$\frac{1}{2}(0^+)$   &  $1425 \pm 50$  & $270 \pm 80$ &  $\pi K$ & $ (93 \pm 10)\% $    \\
   $a_0(1450)$&$1^-(0^{++})$   & $1474 \pm 19$  & $265 \pm 13$ &     $ \left\{\begin{tabular}{@{\ }l@{}}
 $\pi \eta$  \\  $\pi \eta'$ \\ $K \overline{K}$ \end{tabular}\right.$  &  $ \begin{tabular}{@{\ }l@{}}
 $\Gamma(\pi \eta') /\Gamma(\pi \eta)=0.35 \pm 0.16$ \\  $\Gamma(K \overline{K} ) /\Gamma(\pi \eta)=0.88\pm 0.23 $  \end{tabular}$\\
 $\eta(1475)$&$0^+(0^{-+})$  & $1476 \pm 4$  & $85 \pm 9$ &  $ a_0(980)\pi $ \\
 $f_0(1500)$ &$0^+(0^{++})$ & $1505 \pm 6$  & $109 \pm 7$ &     $ \left\{\begin{tabular}{@{\ }l@{}}
 $\pi \pi $ \\  $\pi(1300) \pi$ \\  $\eta \eta$ \\$\eta \eta'$ \\  $K \overline{K}$ \end{tabular}\right.$  &  $ \begin{tabular}{@{\ }l@{}}
 $(34.9 \pm 2.3)\% $ \\  ------   \\  $(5.1 \pm 0.9)\%$ \\ $(1.9 \pm 0.8)\% $ \\  $(8.6 \pm 1.0)\%$ \end{tabular}$\\
  $f_0(1710)$ &$0^+(0^{++})$ & $1720 \pm 6$  & $135 \pm 8$ &     $ \left\{\begin{tabular}{@{\ }l@{}}
 $\pi \pi $  \\  $\eta \eta$ \\ $K \overline{K}$ \end{tabular}\right.$  &  $ \begin{tabular}{@{\ }l@{}}
 $\Gamma(\pi \pi) /\Gamma(K\overline{K})=0.41^{+0.11}_{-0.17}$ \\  $\Gamma(\eta \eta)/\Gamma(K\overline{K})=0.48\pm 0.15 $  \end{tabular}$\\
 $\eta(1760)$&$0^+(0^{-+})$& $1756 \pm 9$& $96 \pm 70$&------  & ------  \\
 \noalign{\hrule height 1pt}
 \noalign{\hrule height 1pt}
\end{tabular}\label{table1}}
\end{table}

\section{Brief review of the generalized linear sigma model}\label{sec2}
\noindent
The effective Lagrangian of GLSM is constructed out of 3$\times$3 matrices $M$ and $M'$ \cite{global}:
\begin{equation}
M = S +i\phi=   \begin{bmatrix}
                   S^1_1+ i\phi^1_1 \qquad & S^2_1+i\phi^2_1 &S^3_1+i\phi^3_1   \\
                    S^1_2+i\phi^1_2   & S^2_2+i \phi^2_2 &S^3_2+i\phi^3_2  \\
                    S^1_3+i\phi^1_3 & S^2_3+i\phi^2_3  & S^3_3+i\phi^3_3
                  \end{bmatrix},
  \\
\hspace{.1cm}
M' = S' +i\phi'=   \begin{bmatrix}
                   S'^1_1+ i\phi'^1_1 & S'^2_1+i\phi'^2_1 &S'^3_1+i\phi'^3_1  \\
                    S'^1_2+i\phi'^1_2   & S'^2_2+i \phi'^2_2 &S'^3_2+i\phi'^3_2  \\
                    S'^1_3+i\phi'^1_3 & S'^2_3+i\phi'^2_3  & S'^3_3+i\phi'^3_3
                  \end{bmatrix},
\label{MMprim}
\end{equation}
where $M$ contains 18 ``bare"
 quark-antiquark scalar and pseudoscalar  fields and
 $M'$  includes 18 ``bare" scalar and pseudoscalar fields
containing two quarks and two antiquarks. Although under chiral transformations
SU(3)$_{\rm L} \times$ SU(3)$_{\rm R}$,  $M $ and $M'$  transform in the same way
\begin{equation}
M \longrightarrow U_L M U^{\dagger}_R, \hskip 2cm
 M' \longrightarrow U_L M' U^{\dagger}_R,
 \end{equation}
 they transform differently under U(1)$_A$
\begin{equation}
 M \longrightarrow e^{2i\nu} M, \hskip 2cm
 M'  \longrightarrow e^{-4i\nu} M'.
  \end{equation}
The Lagrangian density of the model which respects chiral symmetry  SU(3)$_{\rm L} \times$ SU(3)$_{\rm R}$(but not necessarily U(1)$_{\rm A}$ symmetry), is
\begin{equation}
{\cal L} = - \frac{1}{2} {\rm Tr}
\left( \partial_\mu M \partial_\mu M^\dagger
\right) - \frac{1}{2} {\rm Tr}
\left( \partial_\mu M^\prime \partial_\mu M^{\prime \dagger} \right)
- V_0 \left( M, M^\prime \right),
\label{mixingLsMLag}
\end{equation}
where $V_0(M,M^\prime) $ at the leading order $N\leq8$ (eight or fewer underlying quark plus antiquark lines
 at each effective vertex) reads
\begin{eqnarray}
V_0 =&-&c_2 \, {\rm Tr} (MM^{\dagger}) +
c_4^a \, {\rm Tr} (MM^{\dagger}MM^{\dagger})
\nonumber \\
&+& d_2 \,
{\rm Tr} (M^{\prime}M^{\prime\dagger})
     + e_3^a(\epsilon_{abc}\epsilon^{def}M^a_dM^b_eM'^c_f + {\rm H.c.})
\nonumber \\
     &+&  c_3\left[ \gamma_1 {\rm ln} (\frac{{\rm det} M}{{\rm det}
M^{\dagger}})
+(1-\gamma_1){\rm ln}\frac{{\rm Tr}(MM'^\dagger)}{{\rm
Tr}(M'M^\dagger)}\right]^2.
\label{SpecLag}
\end{eqnarray}
     All the terms except the last two are  invariant under  U(1)$_{\rm A}$.
 We  have omitted a possible term $\left[{\rm Tr} (M M^{\dagger})\right]^2$  because it violates the Okubo-Zweig-Iizuka rule.
Also we should add a simple chiral symmetry breaking term due to the small light quark masses to the potential $V_0$
\begin{equation}
V_{SB}=-{\rm Tr}[A(M+M^{\dagger})]=-2 {\rm Tr}[A\, S], \hskip 2cm
A = {\rm diag}(A_1,A_2,A_3),
\end{equation}
where $A_1$, $A_2$ and $A_3$ are proportional to the three light quark masses. The  equilibrium point (or ground state) of the system  can be found by imposing extremum conditions

\begin{equation}\label{mincon}
\left< \frac{\partial V_0}{\partial S}\right>_0 + \left< \frac{\partial V_{SB}}{\partial S}\right>_0=0, \qquad \left< \frac{\partial V_0}{\partial S'}\right>_0 =0,
\end{equation}
where the equilibrium values of fields $S'$, $\phi'$, $S$ and $\phi$  are respectively
\begin{equation}
\langle {S'}_b^a\rangle_0=\delta^b_a \beta_a,   \hskip 1cm
 \langle {\phi'}_a^a\rangle_0=0,   \hskip 1cm
\langle S_b^a \rangle_0=\delta^b_a\alpha_a, \hskip 1cm
 \langle {\phi}_b^a\rangle_0=0.
\end{equation}
\par
In this paper, we assume  isotopic spin
symmetry so that A$_1$ =A$_2\ne$ A$_3$,  $\alpha_1 = \alpha_2  \ne \alpha_3 $  and $\beta_1 = \beta_2  \ne \beta_3.$ The four unknown coefficients in the Lagrangian ($c_2$, $c_4^a$, $d_2$, $e_3^a$)
and the six unknown parameters ($\alpha_1$, $\alpha_3$, $\beta_1$, $\beta_3$, $A_1$, $A_3$) can be determined \cite{global} by using the four  minimum potential conditions (Eq. (\ref{mincon})) together with the following experimental inputs \cite{pdg}
\begin{eqnarray}
 m[a_0(980)] &=& 980 \pm 20\: {\rm MeV},
\nonumber
\\ m[a_0(1450)] &=& 1474 \pm 19\: {\rm MeV},
\nonumber \\
 m[\pi(1300)] &=& 1300 \pm 100\: {\rm MeV},
\nonumber \\
 m_\pi &=& 137 \: {\rm MeV},
\nonumber \\
F_\pi &=& 131 \: {\rm MeV},
\nonumber \\
\frac{A_3} {A_1} &=& 20\rightarrow 30,
\label{inputs1}
\end{eqnarray}
where $A_3/A_1$ is the ratio of strange to nonstrange quark masses \cite{lattice}. Obviously,  there are large uncertainties in the values of $m[\pi(1300)]$ and  $A_3/A_1$  which  dominate the uncertainty of predictions. 
\par
Since the remaining parameters $c_3$ and $\gamma_1$ only affect the isosinglet pseudoscalars, one needs $\eta$ masses as inputs.  From
  Table \ref{exp.}, it is clear that there are two $\eta$'s below  1 GeV which are good candidates for $\eta_1$ and $\eta_2$ predicted by our model
  \begin{eqnarray}
m^{\rm exp.}[\eta (547)] &=& 547.862 \pm
0.018\, {\rm
MeV},\nonumber \\
m^{\rm exp.}[\eta' (958)] &=& 957.78 \pm 0.06
\, {\rm
MeV},
\end{eqnarray}
and four experimental candidates for the two heavier $\eta$'s in our model
 \begin{eqnarray}
m^{\rm exp.}[\eta (1295)] &=& 1294 \pm 4\, {\rm
MeV},\nonumber \\
m^{\rm exp.}[\eta (1405)] &=& 1408.9 \pm 2.4 \,
{\rm
MeV},
\nonumber \\
m^{\rm exp.}[\eta (1475)] &=& 1476 \pm 4\, {\rm
MeV},\nonumber \\
m^{\rm exp.}[\eta (1760)] &=& 1756 \pm 9 \,
{\rm
MeV}.
\end{eqnarray}
This leads to six scenarios to identify the two heavier $\eta$'s ($\eta_3$ and $\eta_4$) with the four experimental candidates above 1 GeV. ${\rm Tr}\left(  M^2_\eta  \right)$ and  ${\rm det} \left(  M^2_\eta  \right)$ are the experimental inputs for the determination of $c_3$ and $\gamma_1$
\begin{eqnarray}
{\rm Tr} \left(  M^2_\eta  \right) &=&
{\rm Tr} \left(  {M^2_\eta}  \right)_{\rm exp.},
\nonumber \\
{\rm det} \left( M^2_\eta \right) &=&
{\rm det} \left( {M^2_\eta} \right)_{\rm exp.}.
\label{trace_det_eq_eta}
\end{eqnarray}
\noindent
For each of these six scenarios,  two sets of $\gamma_1$ and $c_3$  are found as a result of the quadratic  form of $\gamma_1$  in Eq. (\ref{trace_det_eq_eta}). These twelve possibilities are studied in detail in Refs. \citen{global} and \citen{etap} and it was shown that the third scenario with the experimental candidates $\eta(1295)$ and $\eta(1760)$ and solution I (i.e., scenario 3I), has the best agreement with experimental mass spectrum of the $\eta$ system. The variations of these twelve parameters are plotted for different values of  $m[\pi(1300)]$ and $A_3/A_1$ in Ref. \citen{global} and it is shown that the variations of parameters are most affected by the uncertainty in $m[\pi(1300)]$.
\par
After fixing the twelve parameters of the model, we will have the rotation matrices describing the underlying mixing among two- and four-quark components for each spin and isospin state
\begin{equation}
\left[
\begin{array}{cc}
\pi^+(137)\\
{\pi}^+(1300)
\end{array}
\right]
=
R_\pi^{-1}
\left[
\begin{array}{cc}
\phi_1^2\\
{\phi'}_1^2
\end{array}
\right],
\hskip 2cm
\left[
\begin{array}{cc}
K^+(496)\\
{K'}^+(1460)
\end{array}
\right]
=
R_K^{-1}
\left[
\begin{array}{cc}
\phi_1^3\\
{\phi'}_1^3
\end{array}
\right],
\nonumber
\end{equation}

\begin{equation}
\left[
\begin{array}{cc}
a_0^+(980)\\
a_0^+(1450)
\end{array}
\right]
=
L_a^{-1}
\left[
\begin{array}{cc}
S_1^2\\
{S'}_1^2
\end{array}
\right],
\hskip 2cm
\left[
\begin{array}{cc}
K_0(800)\\
K_0^*(1430)
\end{array}
\right]
=
L_\kappa^{-1}
\left[
\begin{array}{cc}
S_1^3\\
{S'}_1^3
\end{array}
\right],
\label{E_Rot1}
\end{equation}
where $R_\pi^{-1}$ and $R_K^{-1}$are the rotation matrices for $I=1$ and $I=1/2$ pseudoscalars and $L_a^{-1}$ and $L_\kappa^{-1}$ are the rotation matrices for $I=1$ and $I=1/2$ scalars. For isosinglet scalars and pseduscalars
\begin{equation}\label{E_Rot2}
\left[
\begin{array}{cc}
f_1\\
f_2\\
f_3\\
f_4
\end{array}
\right]
=
L_0^{-1}
\left[
\begin{array}{cc}
f_a\\
f_b\\
f_c\\
f_d
\end{array}
\right],
\hskip 2cm
 \left[
\begin{array}{cc}
\eta_1\\
\eta_2\\
\eta_3\\
\eta_4
\end{array}
\right]
=
R_{0}^{-1}
\left[
\begin{array}{cc}
\eta_a\\
\eta_b\\
\eta_c\\
\eta_d
\end{array}
\right],
\end{equation}
where $f_i,i=1..4$ and $\eta_i,i=1..4$ are four of the physical isosinglet scalars and pseudoscalars below 2 GeV and
\begin{equation}\label{basis}
 \begin{cases}
 f_a=\frac{S^1_1+S^2_2}{\sqrt{2}} \hskip .87cm
\propto n{\bar n},  \\
 f_b=S^3_3 \hskip 1.7 cm \propto s{\bar s},\\
 f_c=  \frac{S'^1_1+S'^2_2}{\sqrt{2}}
\hskip .7cm \propto ns{\bar n}{\bar s}, \\
f_d= S'^3_3
\hskip 1.5cm \propto nn{\bar n}{\bar n},
 \end{cases}
 \hskip 2cm
  \begin{cases}
\eta_a=\frac{\phi^1_1+\phi^2_2}{\sqrt{2}} \hskip .87cm
\propto n{\bar n}, \\
\eta_b=\phi^3_3 \hskip 1.7cm \propto s{\bar s},\\
\eta_c=  \frac{{\phi '}^1_1 + {\phi'}^2_2}{\sqrt{2}}
\hskip .7cm \propto ns{\bar n}{\bar s},\\
\eta_d= {\phi '}^3_3
\hskip 1.7cm \propto  nn{\bar n}{\bar n}.
 \end{cases}
\end{equation}

In this paper,  we compute two body decay widths of the scalar and psuduescalar mesons using the same order of potential in Ref. \citen{global} with fixed parameters.
This provides  further test of the underlying two and four-quark mixing
among the scalar and pseudoscalar mesons below and above 1 GeV  and the appropriateness of the generalized linear sigma model developed in Ref. \citen{global} and references therein.


\section{Two-body decays}\label{sec3}

\begin{figure}[!htbp]
\begin{center}
\epsfxsize = 3.2 cm
\includegraphics[height=6cm]{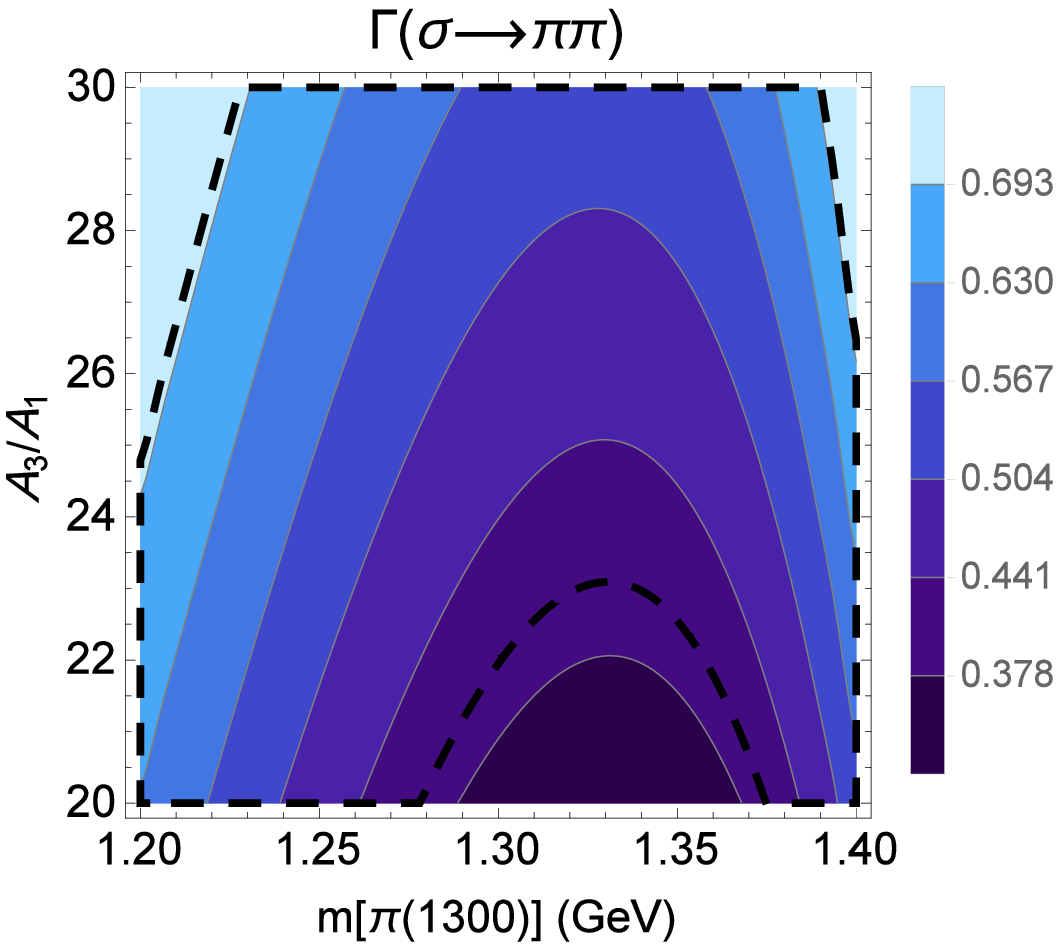}
\hskip .75cm
\epsfxsize = 3.2 cm
\includegraphics[height=6cm]{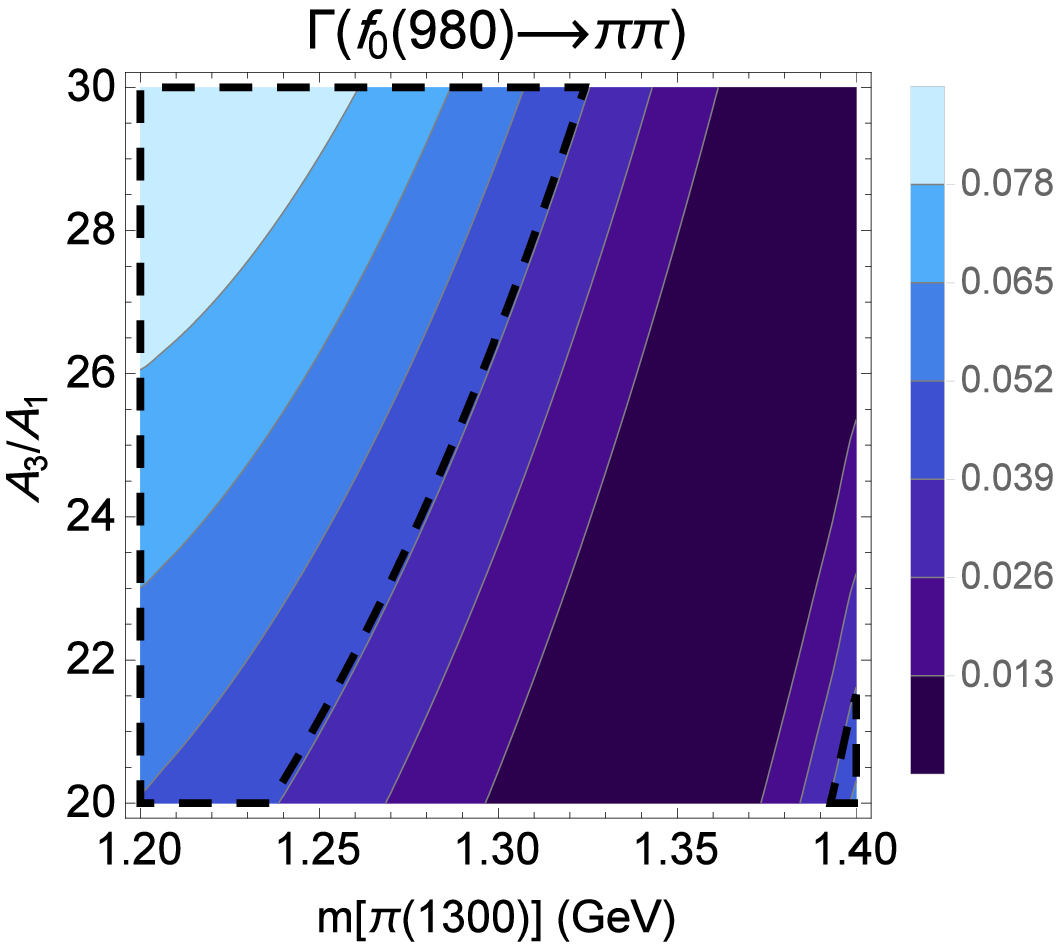}
\vskip 0.75 cm
\epsfxsize = 3.2 cm
\includegraphics[height=6cm]{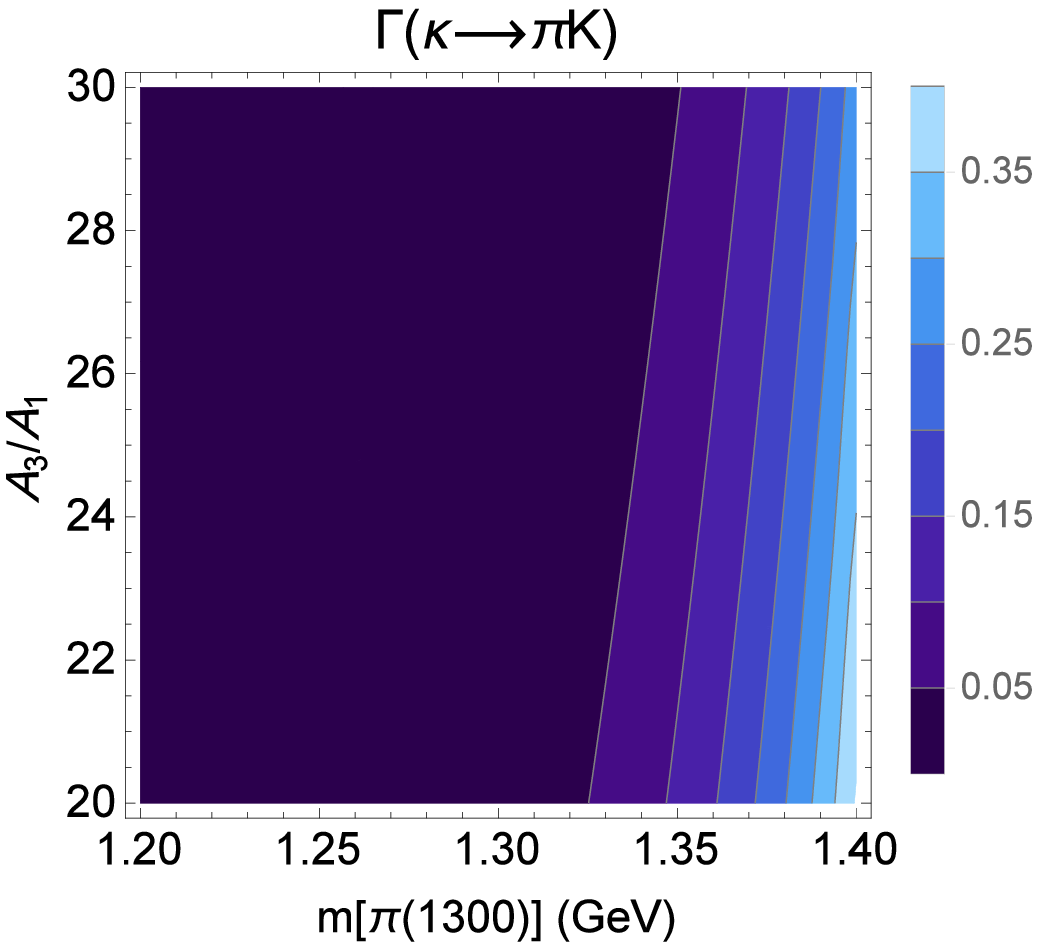}
\hskip .75cm
\epsfxsize = 3.2 cm
\includegraphics[height=6cm]{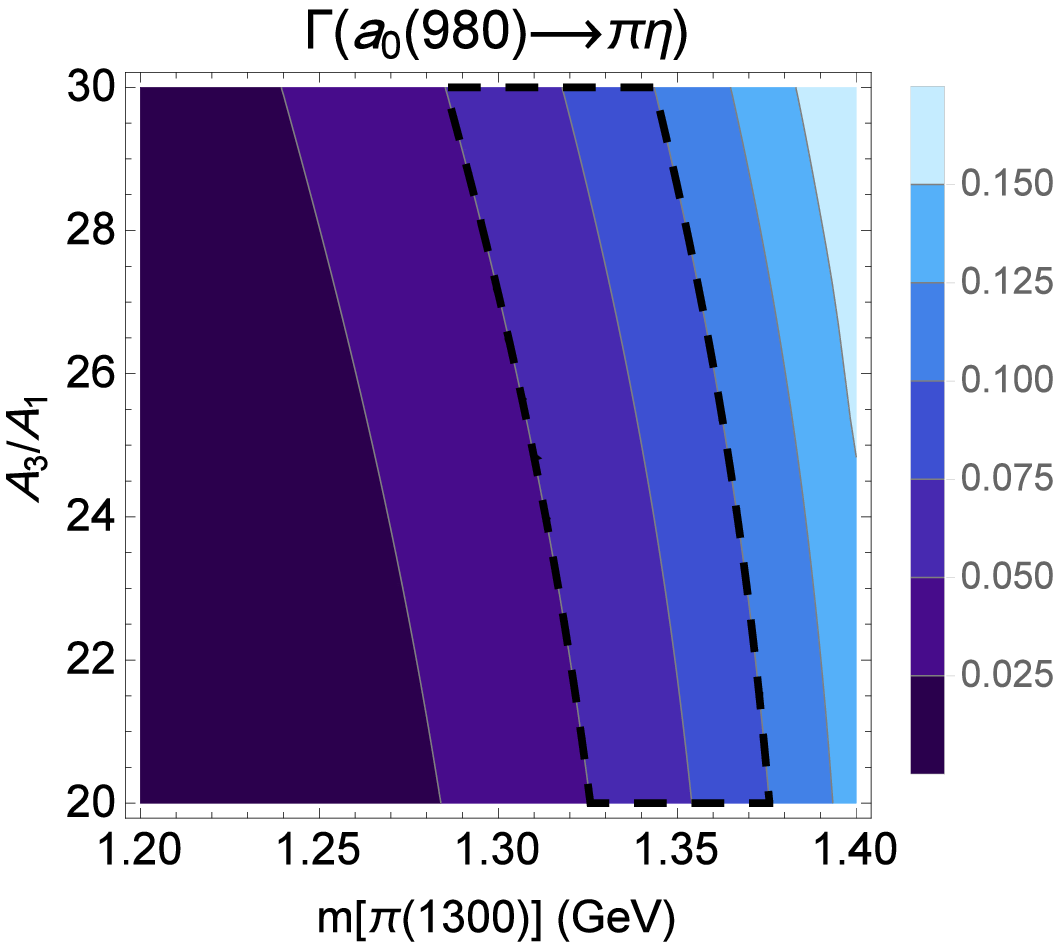}
\caption{Contour plots of the prediction of the model for $\Gamma[\sigma\rightarrow \pi \pi]$ (top left), $\Gamma[f_0(980)\rightarrow \pi \pi]$ (top right), $\Gamma[\kappa\rightarrow \pi K]$ (bottom left) and $\Gamma[a_0(980)\rightarrow \pi \eta]$ (bottom right) over the $m[\pi(1300)]$-$A_3/A_1$ plane. The parameter spaces inside the dashed curves indicate regions for which decay widths are in the experimental range. }
\label{plot1}
\end{center}
\end{figure}

\begin{figure}[!htbp]
\begin{center}
\epsfxsize = 2.7cm
\includegraphics[height=6cm]{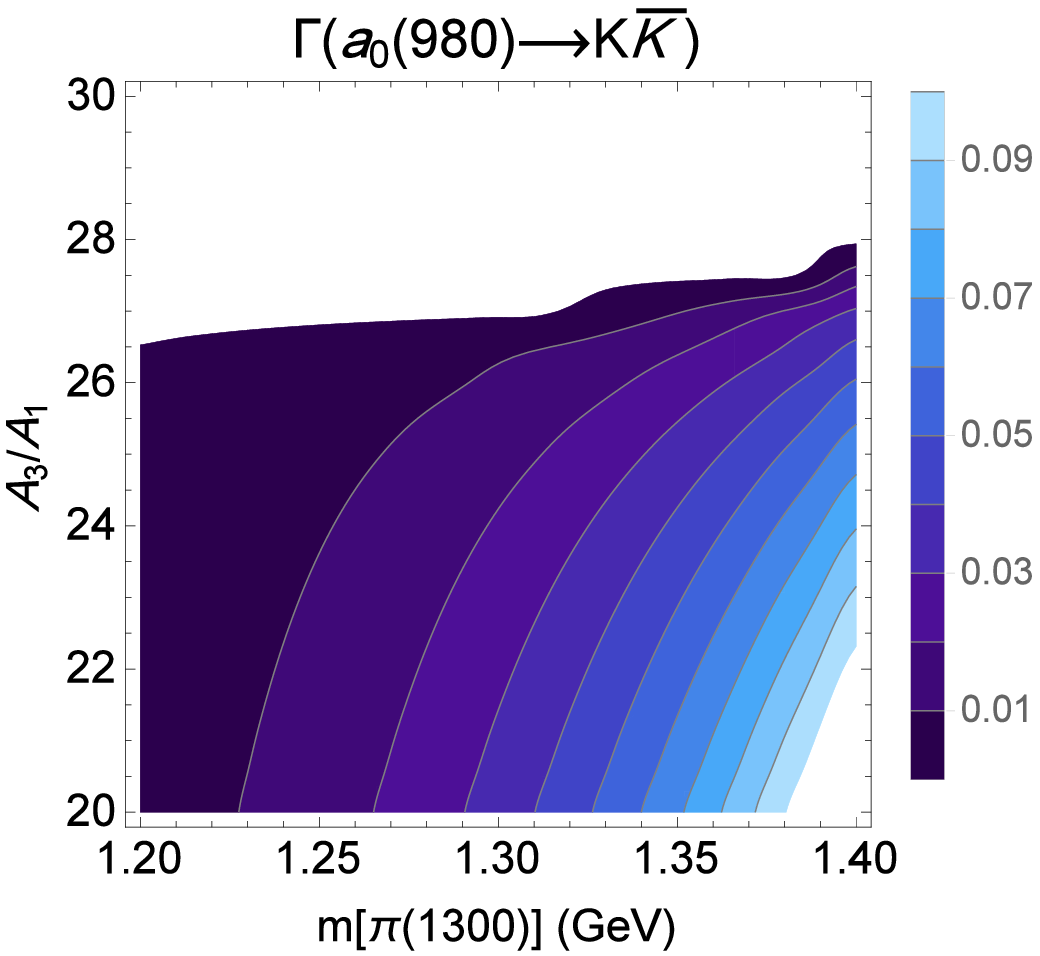}
\hskip 1cm
\epsfxsize = 3.2cm
\includegraphics[height=6cm]{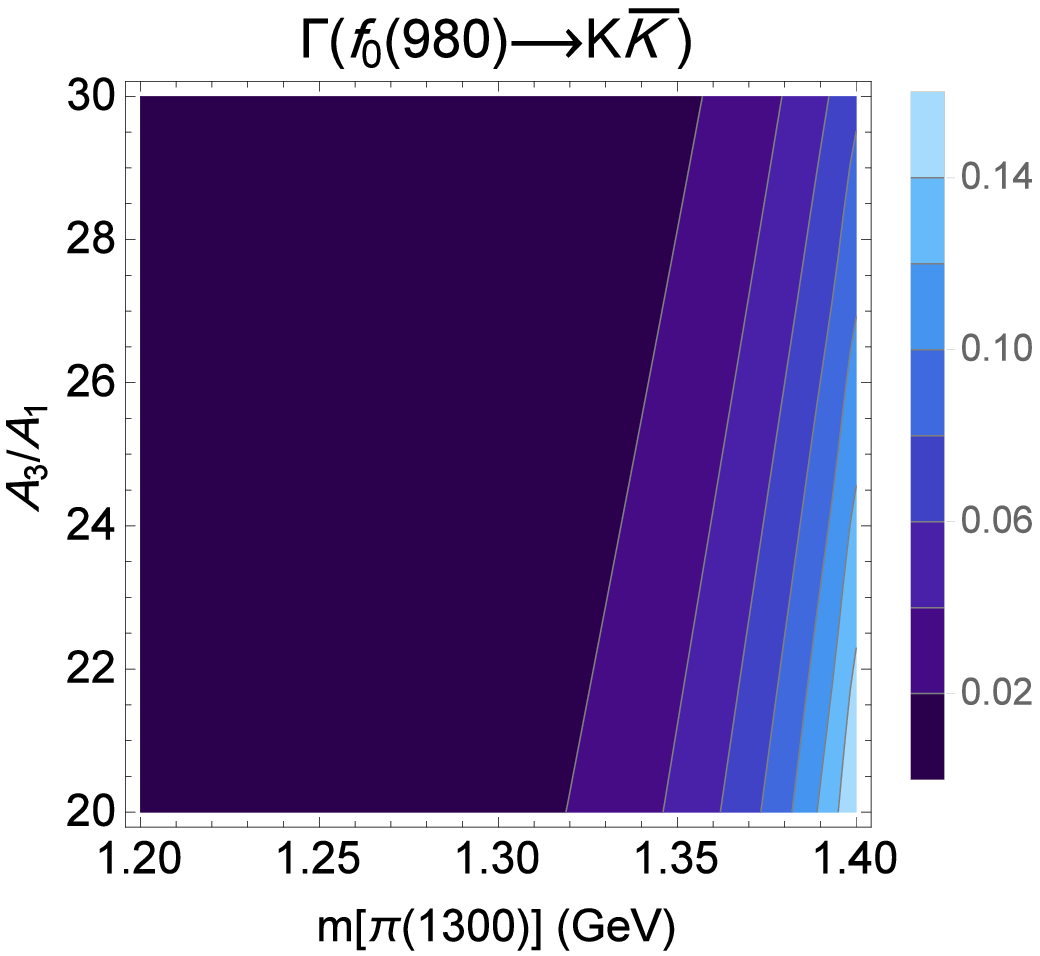}
\caption{Contour plots of the prediction of the model for $\Gamma[a_0(980) \rightarrow K\bar{K}]$ (left) and $\Gamma[f_0(980)\rightarrow K \bar{K}]$ (right) over the $m[\pi(1300)]$-$A_3/A_1$ plane.  The white region belongs to zero value of decay width. As expected, decay widths are small for the main part of the parameter space.}
\label{plot2}
\end{center}
\end{figure}

\begin{figure}[!htbp]
\begin{center}
\epsfxsize = 3.2 cm
\includegraphics[height=6cm]{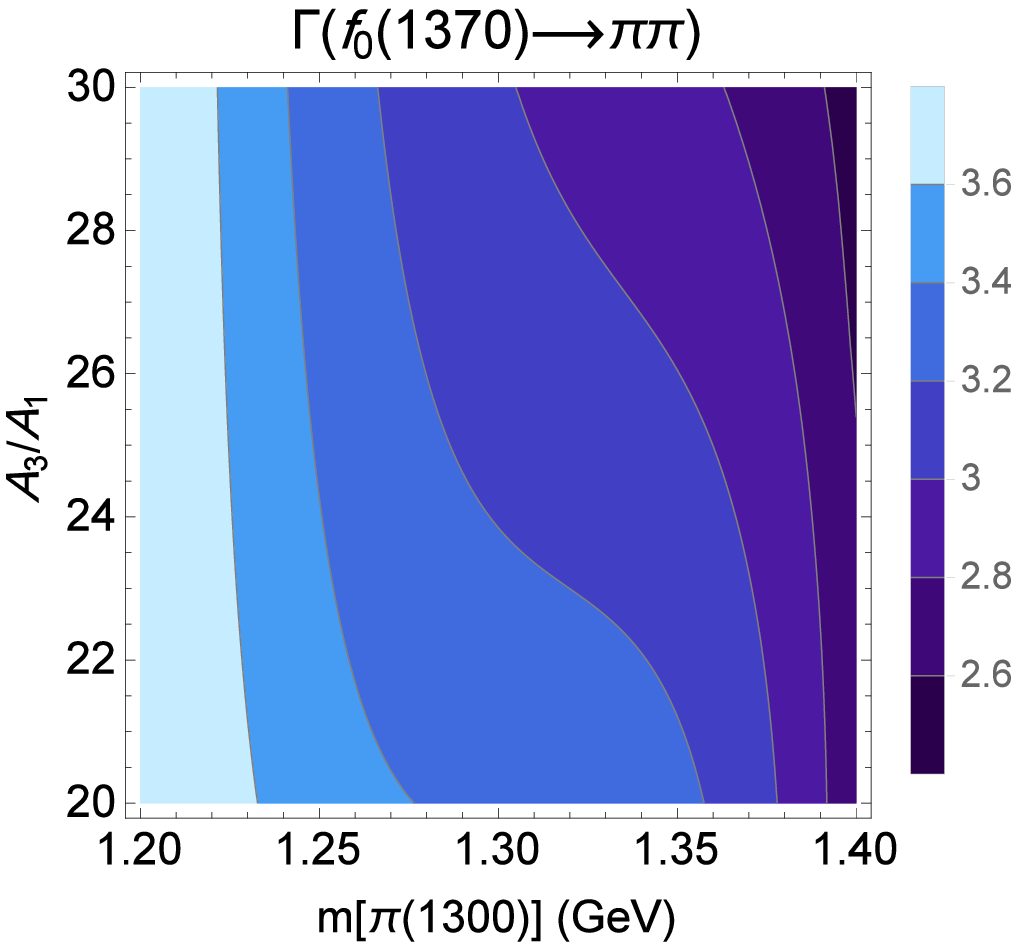}
\hskip 0.5cm
\epsfxsize = 3.2 cm
 \includegraphics[height=6cm]{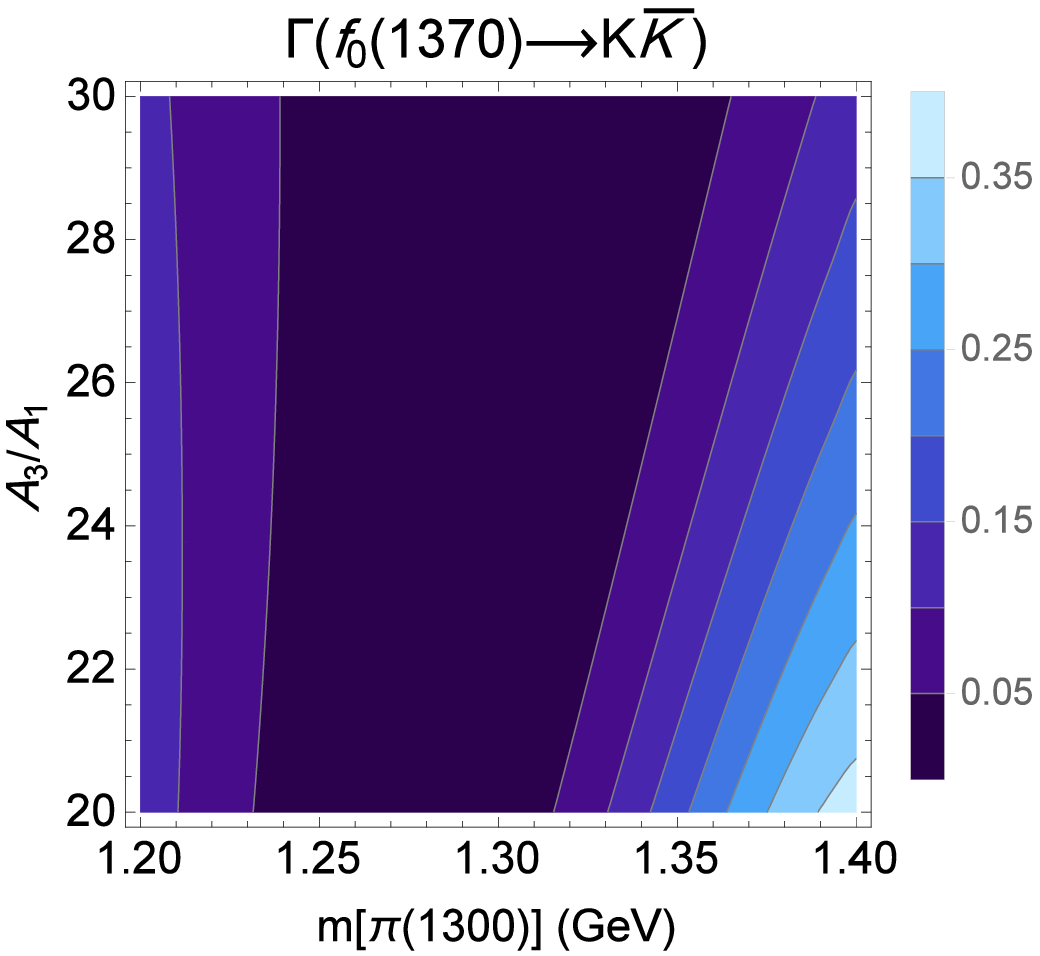}
 \vskip 0.5 cm
 \epsfxsize = 3.2 cm
 \includegraphics[height=6cm]{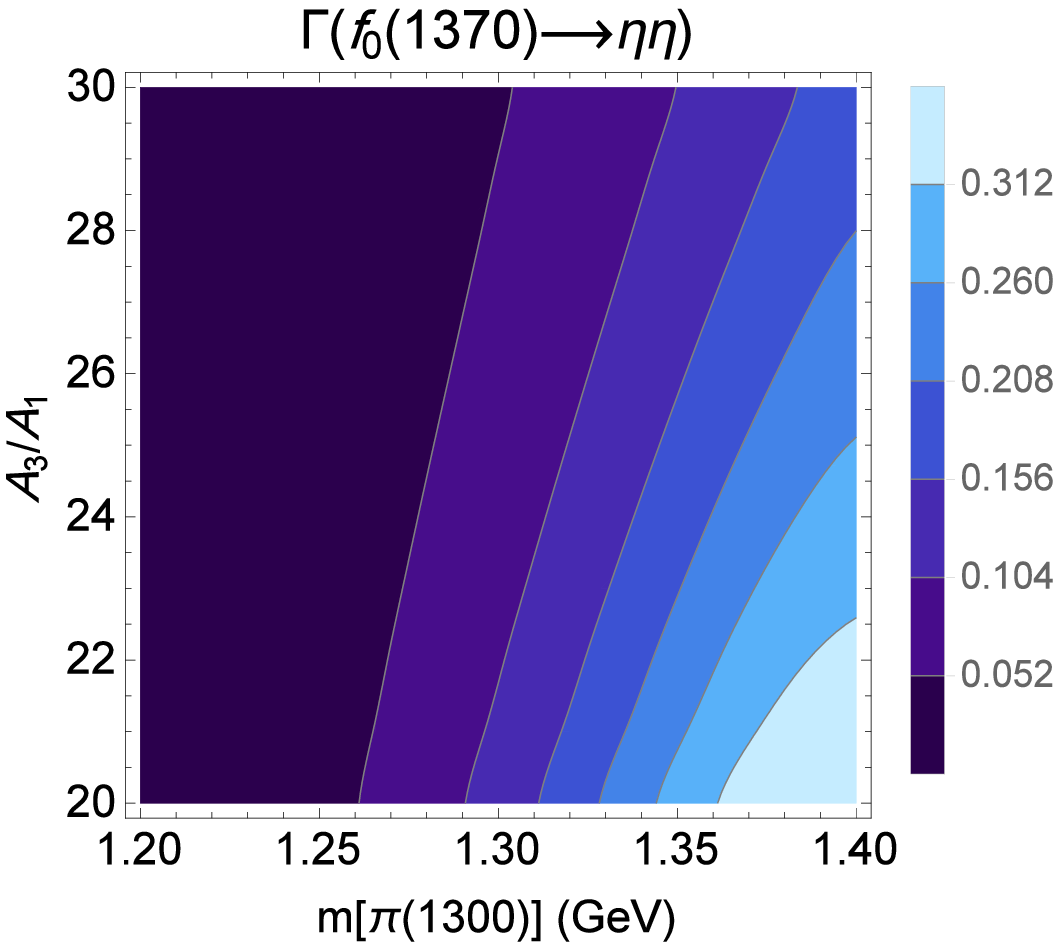}
\hskip 0.5cm
\epsfxsize = 3.2 cm
\includegraphics[height=6cm]{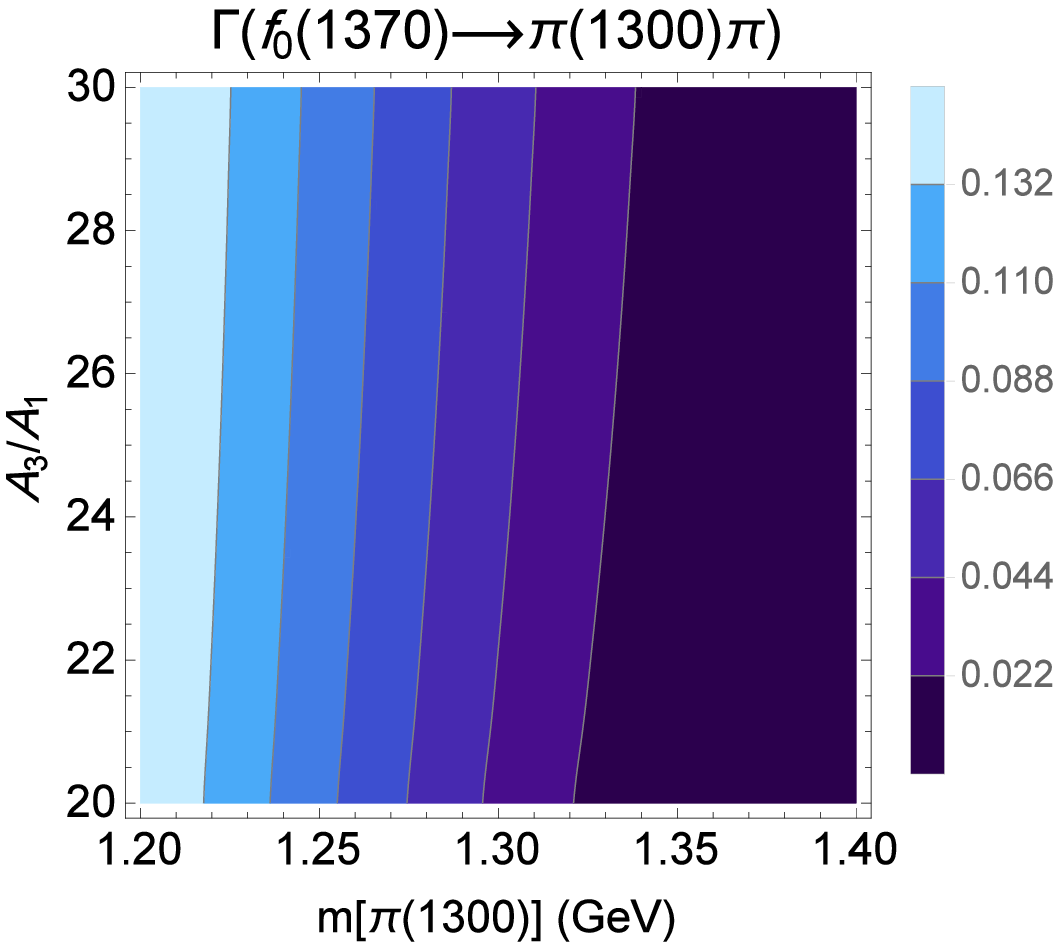}
\caption{Contour plots of the prediction of the model for $\Gamma[f_0(1370)\rightarrow \pi\pi, \, K \bar{K}, \, \eta \eta, \, \pi(1300) \pi]$ over the $m[\pi(1300)]$-$A_3/A_1$ plane. For all decay modes except $\pi \pi$, predicted decay widths are reasonable compared to the full width.}
\label{plot3}
\end{center}
\end{figure}

\begin{figure}[!htbp]
\begin{center}
\epsfxsize = 3.2 cm
\includegraphics[height=6cm]{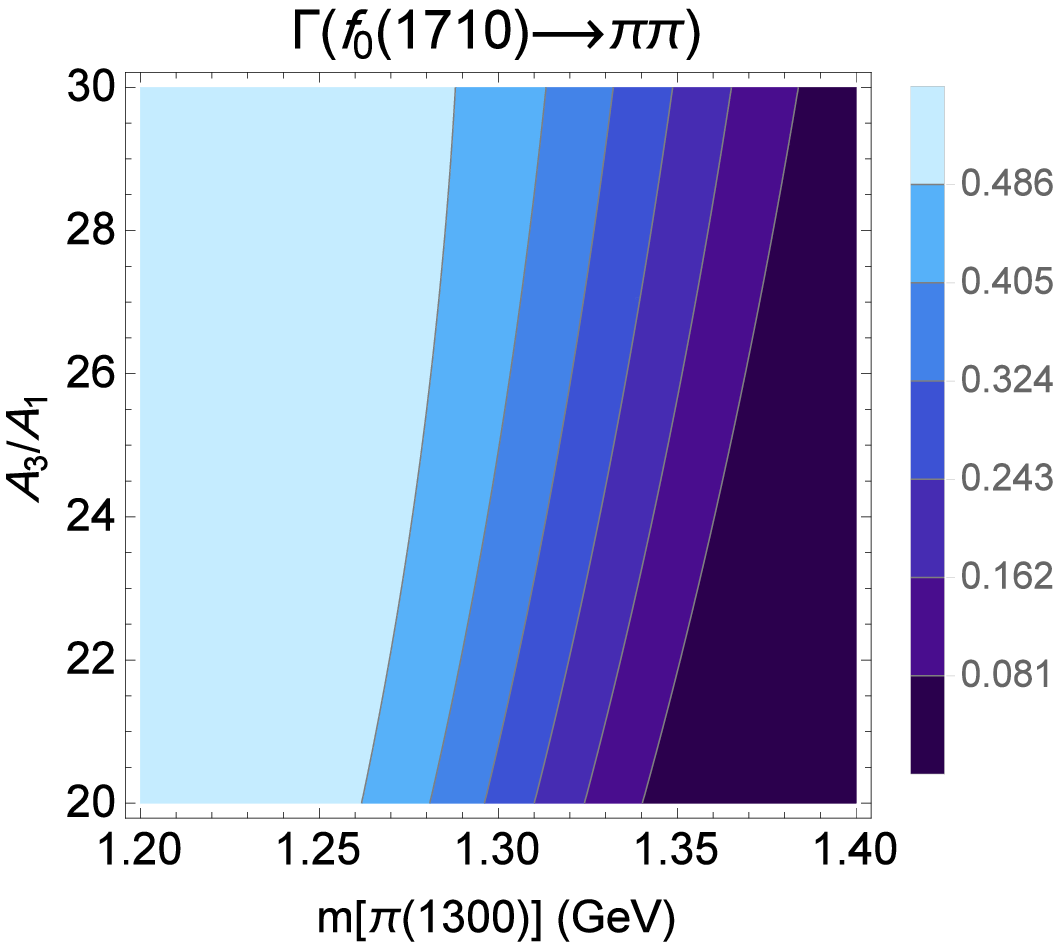}
\hskip 0.5cm
\epsfxsize = 3.2 cm
 \includegraphics[height=6cm]{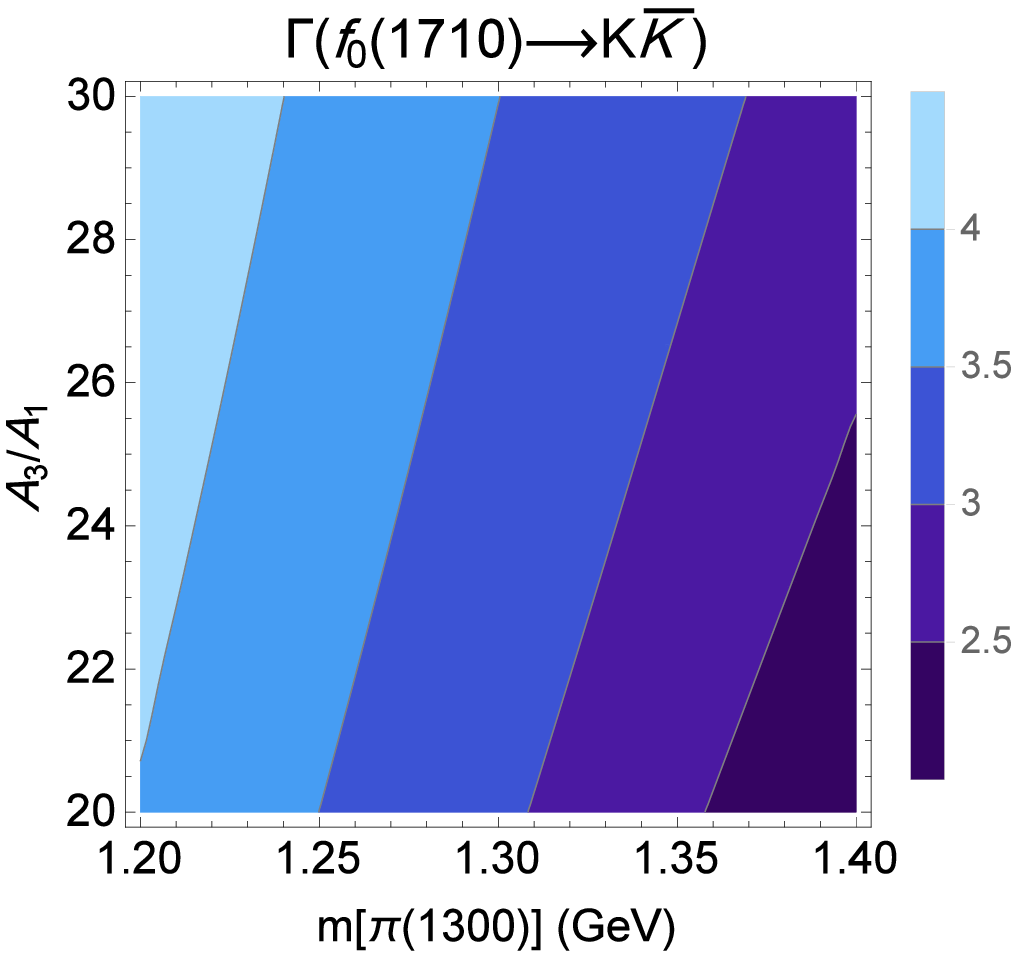}
 \vskip .5cm
\epsfxsize = 3.2 cm
 \includegraphics[height=6cm]{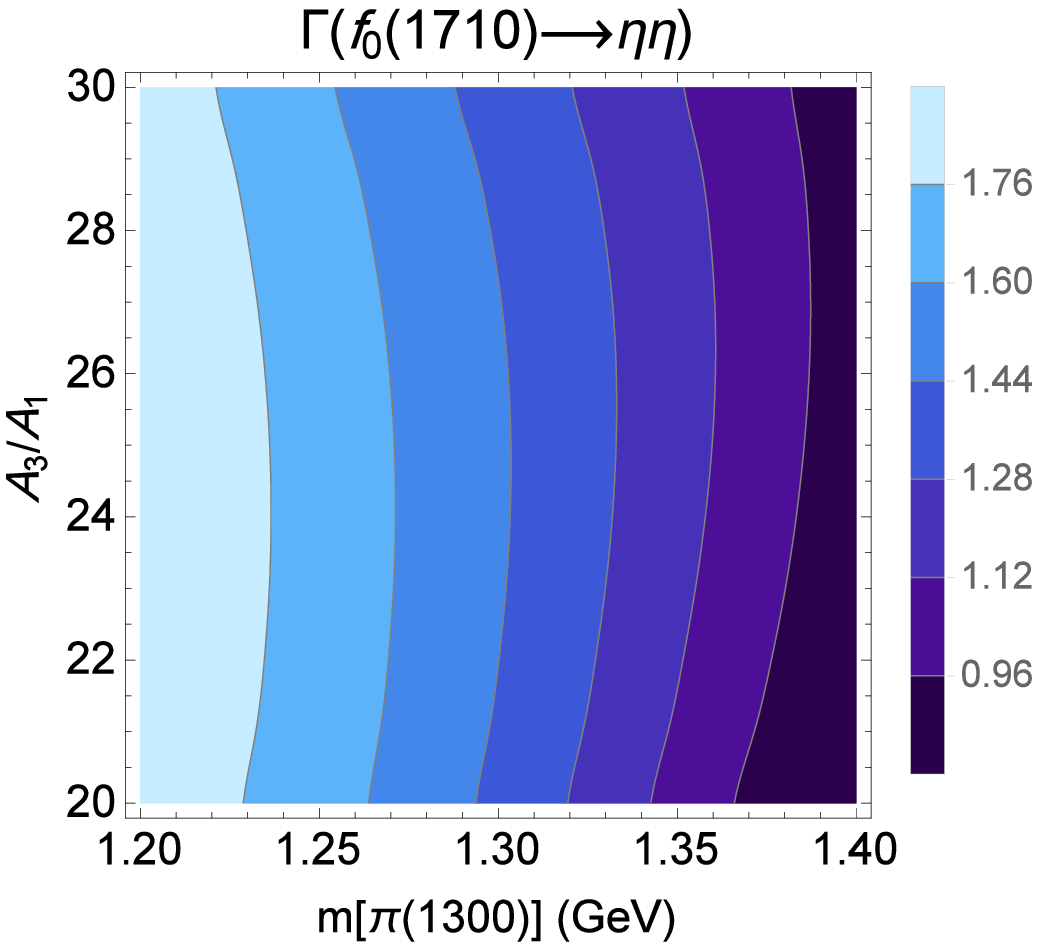}
\caption{Contour plots of the prediction of the model for $ \Gamma[f_0(1710)\rightarrow \pi\pi, \, K\bar{K}, \, \eta \eta] $ decay widths over the $m[\pi(1300)]$-$A_3/A_1$ plane. $K\bar{K}$ and $\eta \eta$ channels have large and unreasonable widths, but for $\pi \pi$ channel, decay width has the right order of magnitude for large values of $m[\pi(1300)]$ compared to the full width. }
\label{plot4}
\end{center}
\end{figure}

\begin{figure}[!htbp]
\begin{center}
\epsfxsize = 3.2 cm
 \vskip .5cm
 \epsfxsize = 3.2 cm
 \includegraphics[height=6cm]{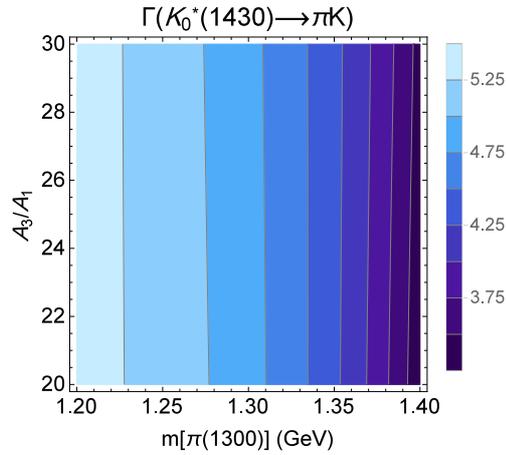}
\caption{Contour plot of the prediction of the model for $K_0^* \rightarrow \pi K  $ decay width over the $m[\pi(1300)]$-$A_3/A_1$ plane. Decay width is too large but receives unitarity corrections due to the $\pi K$ final state interaction.}
\label{plot5}
\end{center}
\end{figure}

\begin{figure}[!htbp]
\begin{center}
\epsfxsize = 3.2 cm
 \includegraphics[height=6cm]{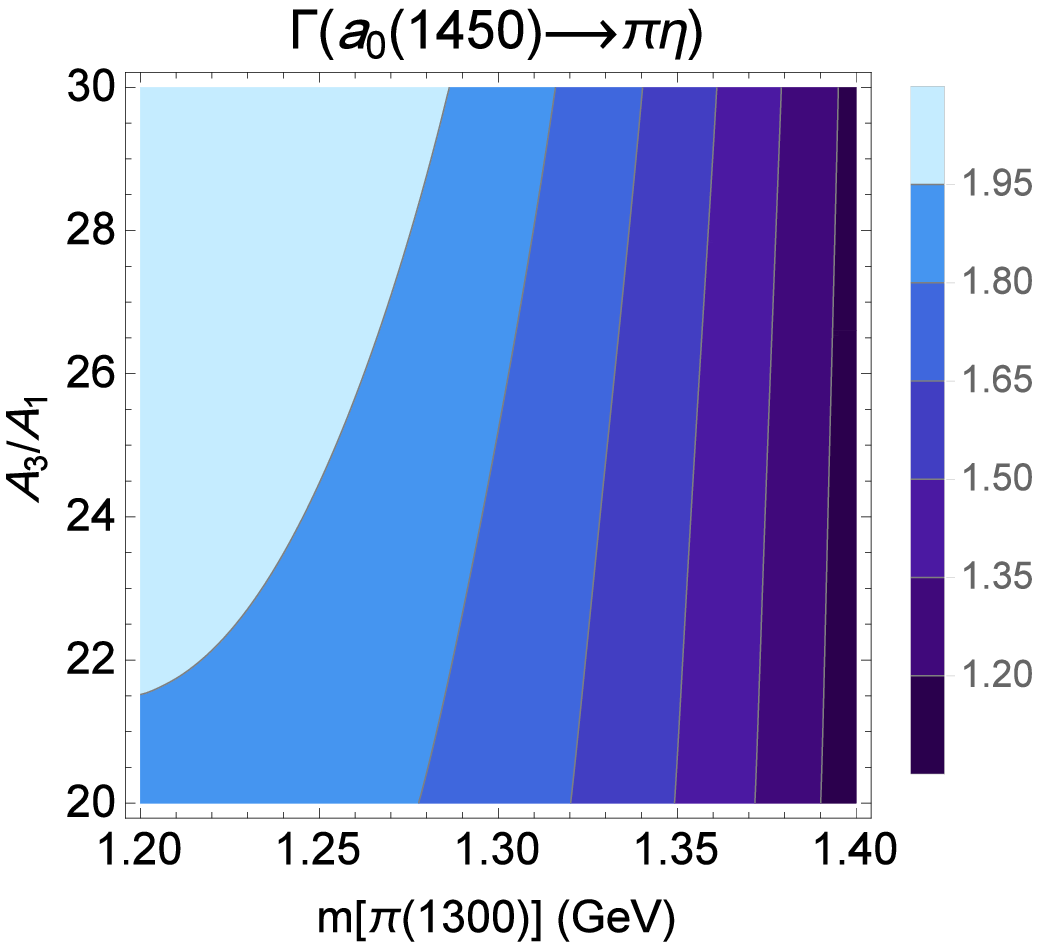}
 \hskip 0.5 cm
 \epsfxsize = 3.2 cm
 \includegraphics[height=6cm]{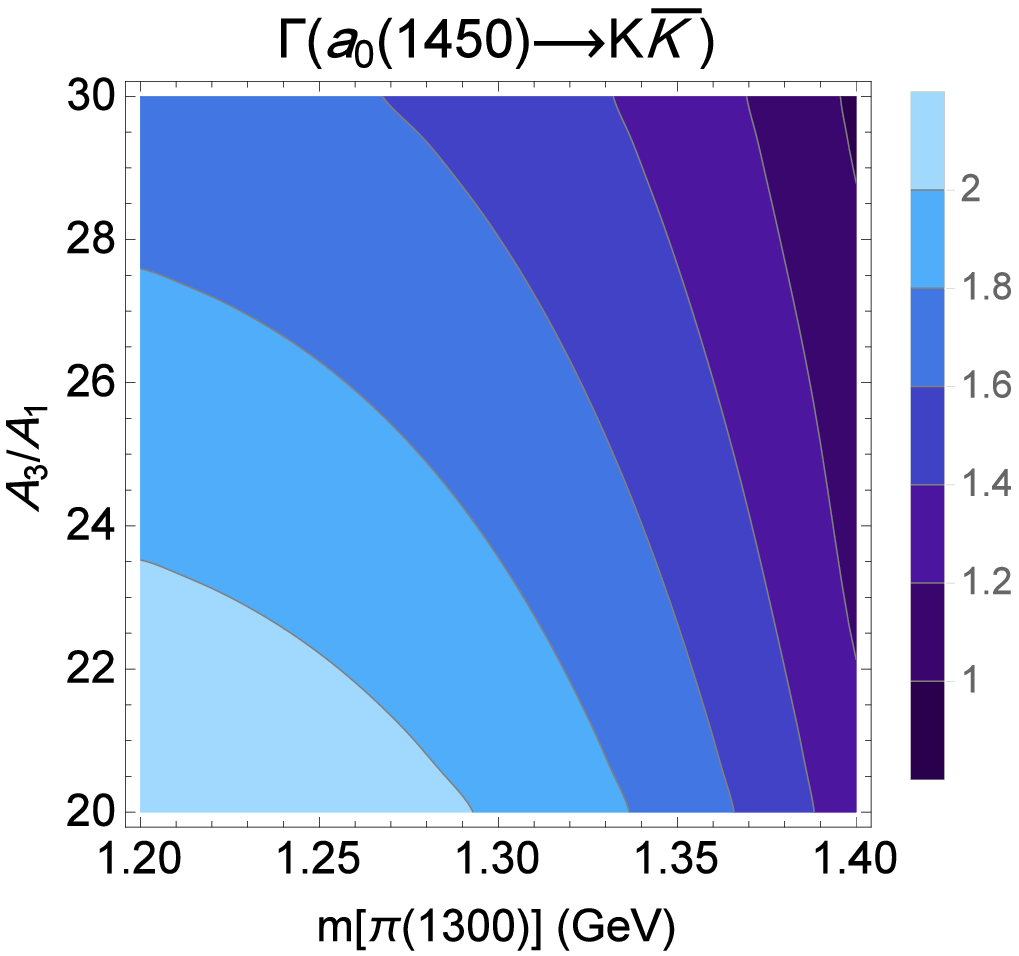}
\vskip 0.5cm
\epsfxsize = 3.2 cm
 \includegraphics[height=6cm]{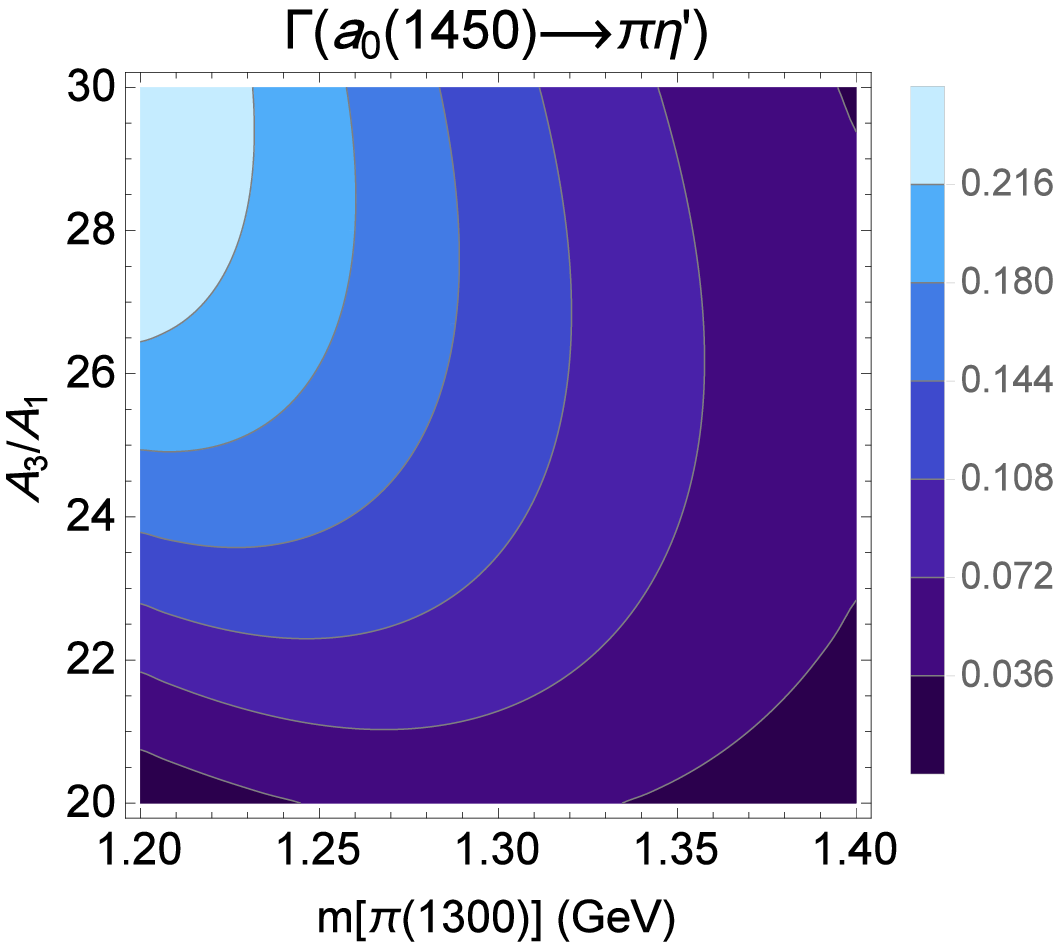}
 \caption{Contour plot of the prediction of the model for $ \Gamma[a_0(1450) \rightarrow \pi \eta, \, K\bar{K}, \, \pi \eta']  $ decay widths over the $m[\pi(1300)]$-$A_3/A_1$ plane. $\Gamma[a_0(1450) \rightarrow \pi \eta']$ is comparable with the experimental full width and the other two decay widths are too large but their ratio is in the experimental range.}
\label{plot6}
\end{center}
\end{figure}

\begin{figure}[!htbp]
\begin{center}
\epsfxsize = 3.2 cm
 \includegraphics[height=6cm]{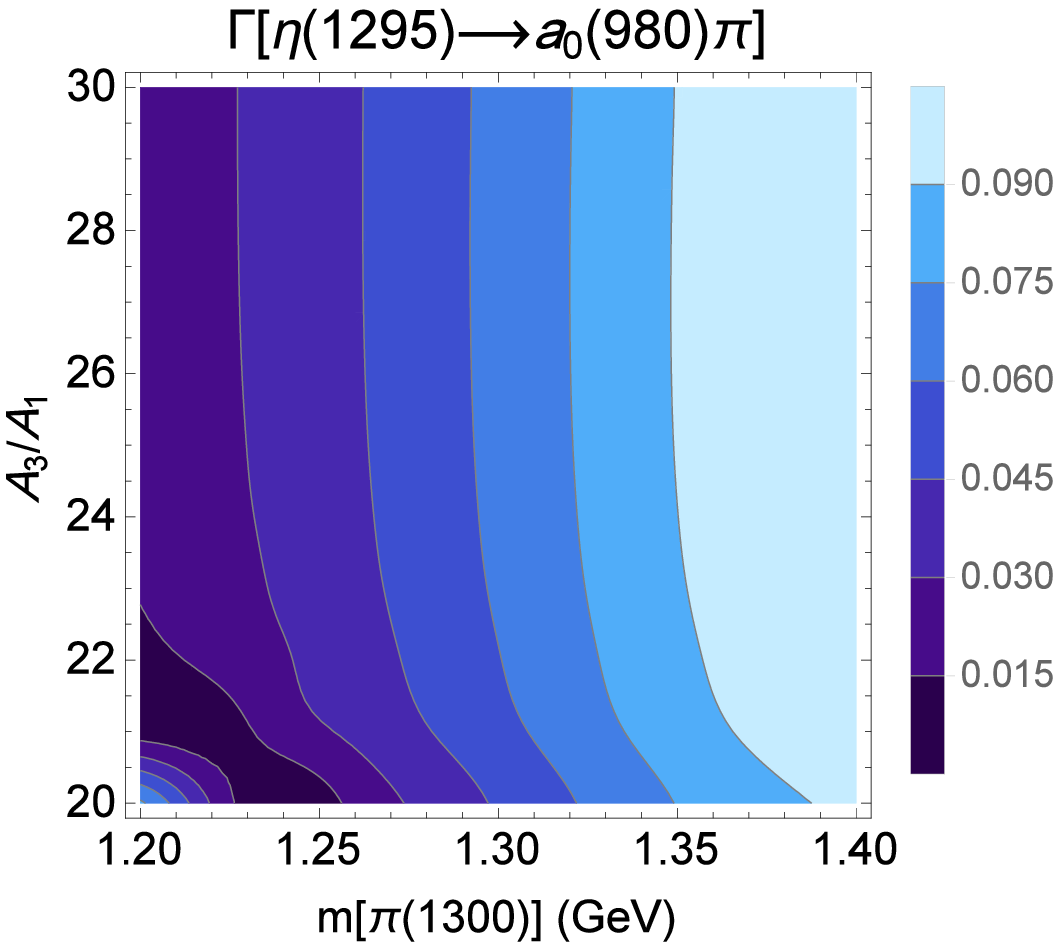}
\hskip 0.5cm
 \epsfxsize = 3.2 cm
 \includegraphics[height=6cm]{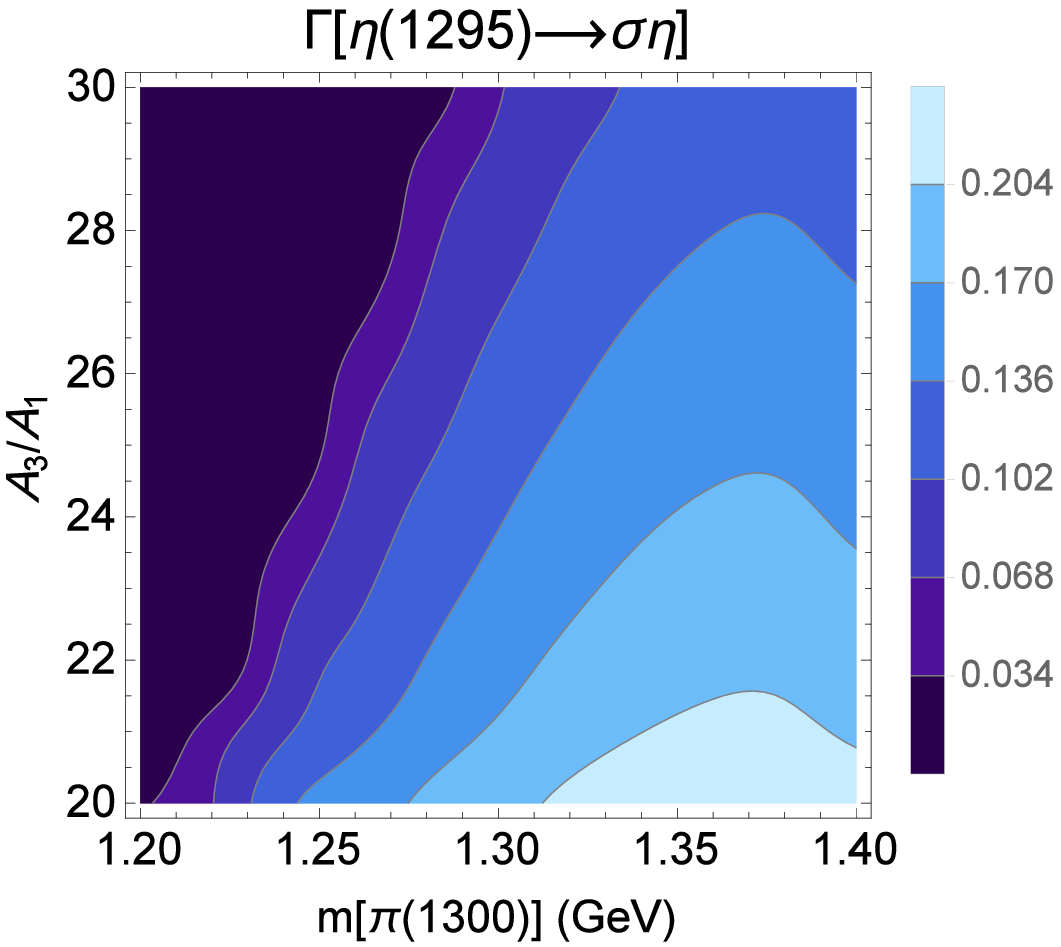}
\caption{Contour plot of the prediction of the model for $\Gamma[\eta(1295)\rightarrow a_0(980)\pi, \, \sigma \eta]$ decay widths over the $m[\pi(1300)]$-$A_3/A_1$ plane. Decay widths are consistent with the experimental data for smaller values of $m[\pi(1300)]$.}
\label{plot7}
\end{center}
\end{figure}

\begin{figure}[!htbp]
\begin{center}
\epsfxsize = 3.2 cm
 \includegraphics[height=6cm]{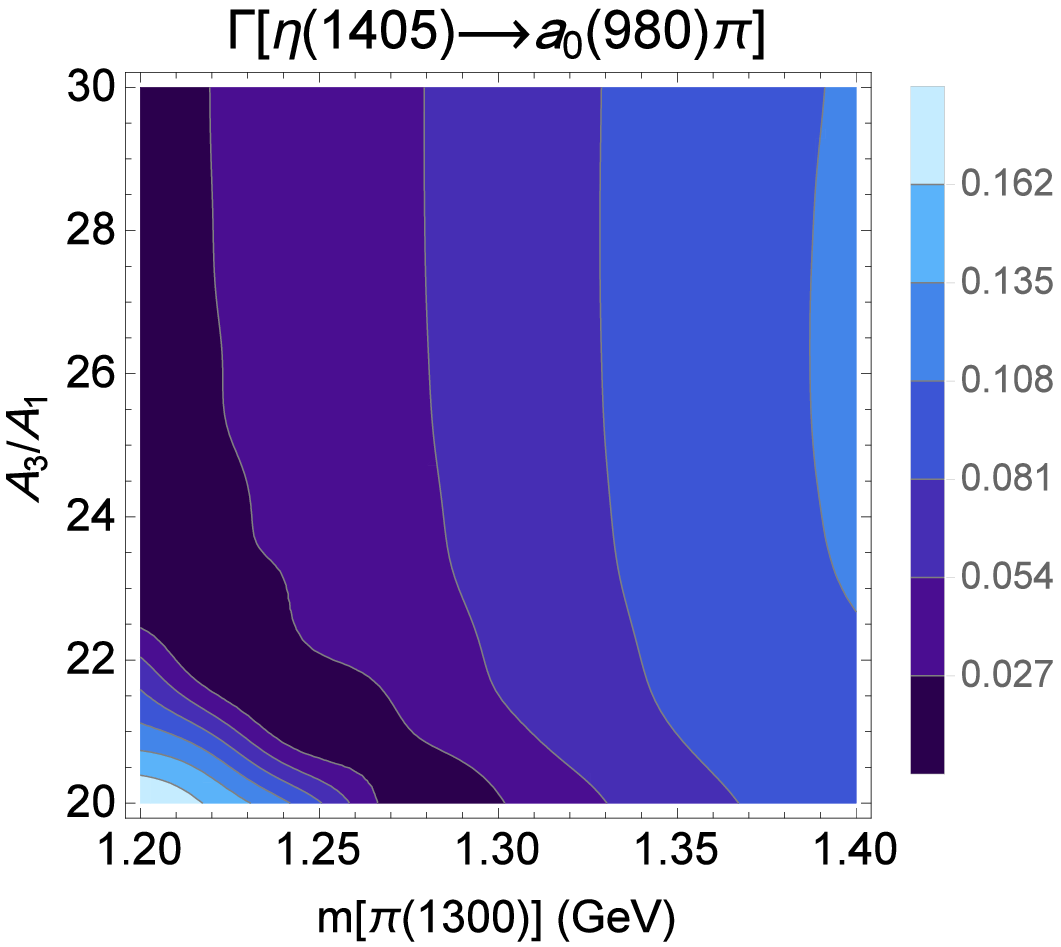}
\hskip .5cm
\epsfxsize = 3.2 cm
\includegraphics[height=6cm]{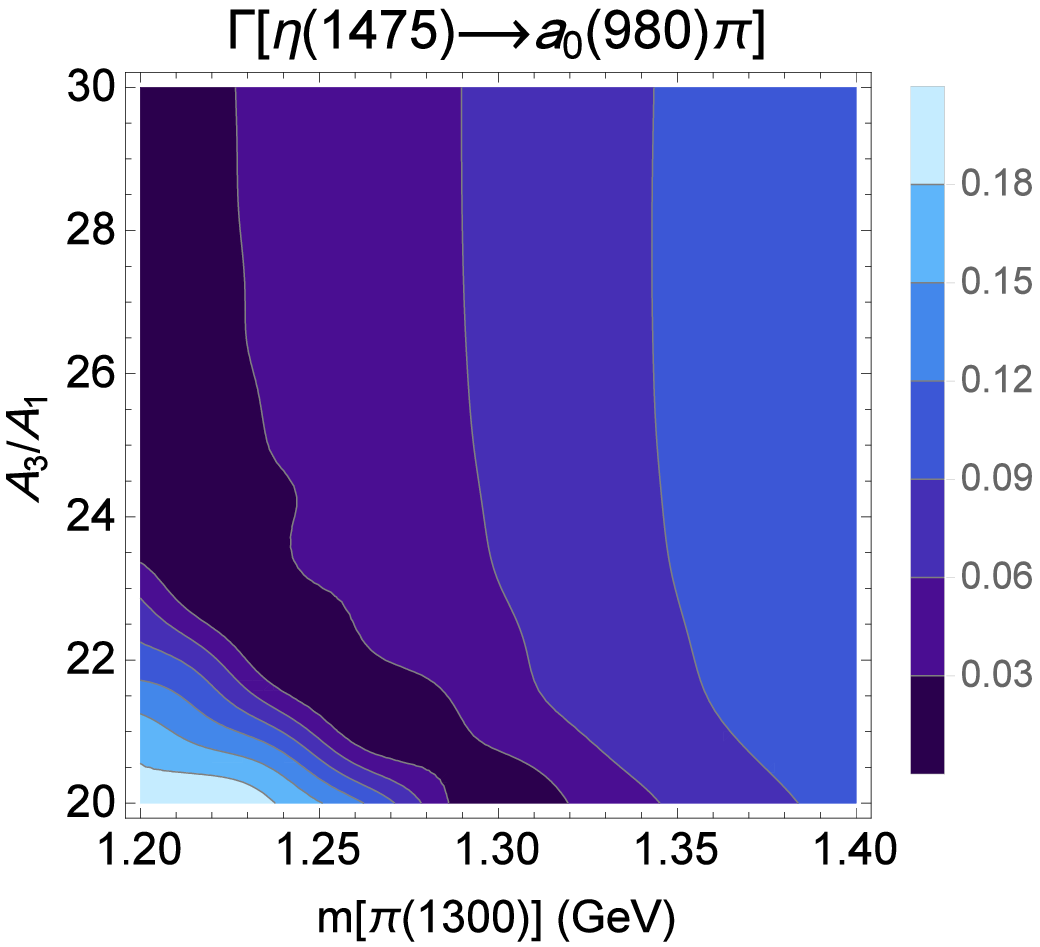}
\caption{Contour plots of the prediction of the model for $\Gamma[\eta(1405)\rightarrow a_0(980)\pi]$ and $ \Gamma[\eta(1475)\rightarrow a_0(980)\pi]$  decay widths over the $m[\pi(1300)]$-$A_3/A_1$ plane. Decay widths are comparable with the full experimental widths. }
\label{plot8}
\end{center}
\end{figure}

\noindent
In this section, we present the prediction of GLSM for hadronic two-body decay widths of lowest lying and next-to-lowest lying scalar and pseudoscalar mesons to show whether the mixing
of scalar and pseudoscalar mesons can improve the results of SNLSM \cite{unitarize} for the lowest lying decays. We also report the
results for the decays above 1 GeV which are beyond the prediction of single nonet model. The decay widths are obtained through the following formulas
\begin{eqnarray}
&&\Gamma[f_i\longrightarrow \pi\pi] = 3\Big(\frac{   q\,\gamma_{f_i\pi\pi}^2 }{8 \pi m_{f_i}^2}\Big), \quad
\Gamma[f_i\longrightarrow K\bar{K}] = \frac{\, q\, \gamma_{f_iKK}^2}{8 \pi m_{f_i}^2},\quad
\Gamma[a_{j}\longrightarrow \pi\eta ] =\frac{\, q\, \gamma_ {a_{j}\pi \eta}^2}{8 \pi m_{a_{j}}^2}, \nonumber\\
&&\Gamma[a_{j}\longrightarrow \pi\eta^{'}] =  \frac{\, q\, \gamma_ {a_{j}\pi \eta^{'}}^2}{8 \pi m_{a_{j}}^2}, \quad
\Gamma[\kappa_l\longrightarrow \pi K]=3\Big( \frac{ q\, \gamma_ {\kappa_l K \pi}^2}{16 \pi m_{\kappa_l}^2}\Big), \quad
\Gamma[a_{j}\longrightarrow K\overline{K}] =  \frac{\, q\, \gamma_ {a_{j}KK}^2}{8 \pi m_{a_{j}}^2},\nonumber\\
 &&\Gamma[f_i\longrightarrow \eta\eta] = \frac{\, q\, \gamma_{f_i\eta\eta}^2}{4 \pi m_{f_i}^2},\quad
\Gamma[f_i\longrightarrow \eta\eta^{'}] = \frac{\, q\, \gamma_{f_i\eta\eta^{'}}^2}{8 \pi m_{f_i}^2}, \quad
\Gamma[f_i\longrightarrow \pi(1300)\pi] = 5\Big(\frac{   q\,\gamma_{f_i\pi\pi'}^2 }{16 \pi m_{f_i}^2}\Big).
\end{eqnarray}
where  $ q$, the center of mass momentum, is given as

\begin{equation}
q=\frac{1}{2m_1}\sqrt{[m_1^2-(m_2+m_3)^2][m_1^2-(m_2-m_3)^2]},
\label{cmq}
\end{equation}
for a general two-body decay $ 1\rightarrow 2\,3 $ and the coupling constants are defined as
\begin{eqnarray}
-{\cal L} &=&
 \frac{\gamma_{f_i \pi \pi}}{\sqrt 2}
f_i \mbox{\boldmath ${\pi}$} \cdot{\mbox{\boldmath ${\pi}$}}+ \frac{\gamma_{f_i K K}}{\sqrt 2}f_i K\overline{K}
+ \frac{\gamma_{a_j KK}}{\sqrt 2} \overline{K} {\mbox{\boldmath ${\tau}$}}  \cdot {\bf a_j} K+ \frac {\gamma_{\kappa_l K \pi}}{\sqrt 2}(\overline{K}{\mbox{\boldmath ${\tau}$}} \cdot {\mbox{\boldmath ${\pi}$}} \kappa_l+\rm{H.c.})\nonumber\\
&&+\,\gamma_{\kappa_l {K} \eta} \left({\overline{ \kappa}_l}  K  {\eta} +\rm{H.c.} \right)+\gamma_{\kappa_l {K} \eta '} \left(
{\bar \kappa_l}  K  {\eta '} + \rm{H.c.} \right)\,+ \gamma_{a_j \pi\eta} {\bf a_j} \cdot  \mbox{\boldmath
${\pi}$} \eta+ \gamma_{a_j \pi\eta'} {\bf a_j} \cdot  \mbox{\boldmath${\pi}$}  \eta'  \nonumber \\
&&+\, \gamma_{f_i \eta \eta} f_i  \eta  \eta+ \gamma_{f_i \eta \eta'} f_i  \eta \eta'+ \gamma_{f_i\eta' \eta'} f_i  \eta'\eta' + \cdots,
\label{lagiso}
\end{eqnarray}
where the subscripts $ i (=1,2,3$ and $ 4 )$, $ j (=1,2) $ and $l(=1,2)$ show the different isosingle, isovector and isodoublet meson states, respectively. These isomultiplets contain the physical fields

\begin{eqnarray}\label{isomul}
&&K=
\left[
\begin{array}{cc}
K^+\\
K^0
\end{array}
\right],\qquad
\overline{K}=
\left[
K^-\; \overline{K}^0
\right],\qquad
\kappa=
\left[
\begin{array}{cc}
\kappa^+\\
\kappa^0
\end{array}
\right],\qquad
\overline{\kappa}=
\left[
\kappa^-\; \overline{\kappa}^0
\right],\nonumber\\
&&\pi_1=\frac{1}{\sqrt{2}}(\pi^+ +\pi^-),\qquad \pi_2=\frac{i}{\sqrt{2}}(\pi^+ - \pi^-),\qquad \pi_3=\pi^0, \nonumber\\
&&a_{01}=\frac{1}{\sqrt{2}}(a_0^+ +a_0^-),\qquad a_{02}=\frac{i}{\sqrt{2}}(a_0^+ - a_0^-),\qquad a_{03}=a_0^0.
\end{eqnarray}
The coupling constants are related to the bare couplings through the following relations
\begin{align*}
\gamma_{f_i\pi\pi}&=
{1\over {\sqrt{2}}}
\left\langle
{{\partial^3 V}
\over
{\partial f_i \, \partial \pi^+ \, \partial \pi^-}}
\right\rangle
=
{1\over {\sqrt{2}}}
\sum_{I,A,B}
\left\langle
{{\partial^3 V}
\over
{
 \partial f_I \,
 \partial (\phi_1^2)_A \,
 \partial (\phi_2^1)_B
}}
\right\rangle
(L_0)_{I i} \,
(R_\pi)_{A1} \,
(R_\pi)_{B1},
\nonumber \\
\gamma_{f_i\pi\pi'} &=
{1\over {\sqrt{2}}}
\left\langle
{{\partial^3 V}
\over
{\partial f_i \, \partial \pi^+ \, \partial \pi^-}}
\right\rangle
=
{1\over {\sqrt{2}}}
\sum_{I,A,B}
\left\langle
{{\partial^3 V}
\over
{
 \partial f_I \,
 \partial (\phi_1^2)_A \,
 \partial (\phi_2^1)_B
}}
\right\rangle
(L_0)_{I i} \,
(R_\pi)_{A1} \,
(R_\pi)_{B2},
\nonumber \\
\gamma_{f_i K K} &=
\sqrt 2
\left\langle
{{\partial^3 V}
\over
{\partial f_i \, \partial K^+ \, \partial K^-}}
\right\rangle
=
\sqrt 2
\sum_{I,A,B}
\left\langle
{{\partial^3 V}
\over
{
 \partial f_I \,
 \partial (\phi_1^3)_A \,
 \partial (\phi_3^1)_B
}}
\right\rangle
(L_0)_{I i} \,
(R_K)_{A1} \,
(R_K)_{B1},
\nonumber \\
\gamma_{a_j\pi\eta} &=
\left\langle
{{\partial^3 V}
\over
{\partial a_j^- \, \partial \pi^+ \, \partial \eta}}
\right\rangle
=
\sum_{A,B,I}
\left\langle
{{\partial^3 V}
\over
{
 \partial (S^2_1)_A \,
 \partial (\phi_1^2)_B \,
 \partial \eta_I
 }}
\right\rangle
(L_a)_{Aj} \,
(R_\pi)_{B1} \,
(R_0)_{I1},
\nonumber \\
\gamma_{a_j\pi\eta'} &=
\left\langle
{{\partial^3 V}
\over
{\partial a_j^- \, \partial \pi^+ \, \partial \eta}}
\right\rangle
=
\sum_{A,B,I}
\left\langle
{{\partial^3 V}
\over
{
 \partial (S^2_1)_A \,
 \partial (\phi_1^2)_B \,
 \partial \eta_I
 }}
\right\rangle
(L_a)_{Aj} \,
(R_\pi)_{B1} \,
(R_0)_{I2},
\nonumber \\
\gamma_{\kappa_l K \pi} &=
\left\langle
{{\partial^3 V}
\over
{\partial \kappa_l^0 \, \partial K^- \, \partial \pi^+}}
\right\rangle
=
\sum_{A,B,C}
\left\langle
{{\partial^3 V}
\over
{
\partial (S_2^3)_A \,
\partial  (\phi_3^1)_B \,
\partial (\phi_1^2)_C
 }}
\right\rangle
(L_\kappa)_{Al} \,
(R_K)_{B1} \,
(R_\pi)_{C1},
 \nonumber \\
\gamma_{a_j K K} &=
\left\langle
{{\partial^3 V}
\over
{\partial a_j^+ \, \partial K^0 \, \partial K^-}}
\right\rangle
=
\sum_{A,B,C}
\left\langle
{{\partial^3 V}
\over
{
\partial (S_1^2)_A \,
\partial (\phi_2^3)_B \,
\partial (\phi_3^1)_C
}}
\right\rangle
(L_a)_{A j} \,
(R_K)_{B1} \,
(R_K)_{C1},
\nonumber \\
 \gamma_{f_i\eta\eta}&=
 \frac{1}{2}\left\langle
{{\partial^3 V}
\over
{\partial f_i \, \partial \eta \, \partial \eta}}
\right\rangle=
 \frac{1}{2}\sum_{I,J,K}\left\langle\frac{\partial^{3}V}{\partial f_{I}\partial\eta_{J}\partial\eta_{K}}\right\rangle (L_{0})_{Ii}(R_{0})_{J1}(R_{0})_{K1},
  \nonumber \\
   \gamma_{f_i\eta\eta'}&=
\left\langle
{{\partial^3 V}
\over
{\partial f_i \, \partial \eta \, \partial \eta'}}
\right\rangle=
\sum_{I,J,K}\left\langle\frac{\partial^{3}V}{\partial f_{I}\partial\eta_{J}\partial\eta_{K}}\right\rangle (L_{0})_{Ii}(R_{0})_{J1}(R_{0})_{K2}.\numberthis
\end{align*}
where $ A $, $ B $, $ C =1, 2$ with $ 1 $
denoting nonet $ M $ and $ 2 $ denoting nonet $ M' $ and
$ I $, $ J $, $ K =a, \,b,\, c $ and $d$ represent the four bases in Eq. (\ref{basis}). $ L_0 $,
$ R_\pi $, $ L_a $, $ R_0 $, $ L_\kappa $, and $ R_K $ are the rotation matrices
defined in Eqs. (\ref{E_Rot1}) and (\ref{E_Rot2}). The nonvanishing bare three-point coupling constants are given in Appendix A.  The contour plots of all the hadronic two-body decay widths below 2 GeV are presented in Figs. \ref{plot1}-\ref{plot8}. In the following we will explain each figure separately:\\ \\

\noindent{\underline{\textbf{Figure \ref{plot1}}}}:\\
\par
In this figure we review the results for $\sigma\rightarrow \pi \pi$, $f_0(980) \rightarrow\pi \pi$, $a_0(980)\rightarrow \pi \eta$ and $ \kappa \rightarrow \pi K$ decay widths which are given in Ref. \citen{etap}.
Contour plot for $\sigma\rightarrow \pi \pi$ decay width shows that for a large part of the parameter space, the prediction of the model for this decay overlaps  with the experimental data (The region inside the dashed curves show the experimental range for each decay).
Also for $f_0(980) \rightarrow\pi \pi$ and $a_0(980)\rightarrow \pi \eta$ decay widths, there exist regions of parameter space which agree with the experimental ranges. No region coincides with the experimental range for $ \kappa \rightarrow \pi K$ decay width, but for $m[\pi(1300)]$ near $1.4$ GeV, decay width reaches $400$ MeV which at least has a right order of magnitude compared to PDG data. The interesting point is that due to final state interactions in $\pi K$ scattering,  total decay width for $\kappa$  gets larger (It is shown in Ref. \citen{piK} and will be discussed in the next section).\\ \\

\noindent{\underline{\textbf{Figure \ref{plot2}}}}:\\
\par
Since $\pi \pi$ and $\pi \eta$ decay modes are dominant channels for decays of $f_0(980)$ and $a_0(980)$, respectively,
  $K\bar{K}$ decay channel for these two particles should have narrow decay width. Fig. \ref{plot2} shows that the model prediction for these
 channels fulfill our expectation: decay widths are small for some regions of the parameter space.  Averaging over the entire parameter space, we have
 \begin{eqnarray}
\Gamma[a_0(980)\rightarrow K\bar{K}]&=&21 \pm 27,\nonumber\\
\Gamma[f_0(980)\rightarrow K\bar{K}]&=&21 \pm 31,
 \end{eqnarray}
 where the uncertainty is the standard deviation of the predicted data. These values are in agreement with the results of Refs. \citen{tb1} and \citen{tb2}.
 The prediction of the model over the $m[\pi(1300)]-A_3/A_1$ plane for $  \Gamma[a_0(980)\rightarrow K\bar{K}]/\Gamma[a_0(980)\rightarrow \pi \eta]$ is
 \begin{equation}
 \frac{\Gamma[a_0(980)\rightarrow K\bar{K}]}{\Gamma[a_0(980)\rightarrow \pi \eta]}=0.44\pm0.41.
 \end{equation}
 The mean value is about $2.5$ times larger than the experimental value (Table \ref{table1}) but the uncertainty covers the observed range.\\ \\
\noindent{\underline{\textbf{Figure \ref{plot3}}}}:\\
\par
Now let us study hadronic decay widths above 1 GeV.
It is shown in Ref.  \citen{global} that the GLSM prediction for the third f-meson mass is $1504 \pm 6$, but due to unitarity corrections in $\pi\pi$ scattering \cite{pipi}, its mass reduces to $1149\pm43$ which is close to $f_0(1370)$ mass. Furthermore, we know that broad states receive  more  contributions (in mass and width) from the unitarity corrections compared to the narrow ones.  Therefore we identify the third f-meson predicted by our model with the broad state $f_0(1370)$ and not the narrow state $f_0(1500)$. This is consistent with the result of Ref. \citen{glueball} in which it was shown that  $f_0(1500)$ is the scalar glueball state.
 Moreover,  $f_0(1370)$  was shown to be predominantly a quark-antiquark state (particularly $s\bar{s}$) with small remnant of four quark and glue components \cite{mix8}. Although there is no experimental data for decay widths of different modes of this broad resonance, Fig. \ref{plot3} shows that for all the decay modes except $\pi \pi$, the predicted decay widths are reasonable compared to the full width but $\Gamma[f_0(1370)\rightarrow \pi \pi]$  is extremely large and far from experiment. The predicted averaged value for $\Gamma[f_0(1370) \rightarrow K \bar{K}]$ is $72 \pm 76 $  MeV in agreement with Ref. \citen{tb3}.\\ \\
\noindent{ \underline{\textbf{Figure \ref{plot4}}}}:\\
\par
 For $f_0(1710)$ which decays predominantly into kaons, $K\bar{K}$ and $\eta \eta$ channels have large and unreasonable widths but their ratio is
\begin{equation}
 \frac{\Gamma[f_0(1710)\rightarrow \eta \eta]}{\Gamma[f_0(1710)\rightarrow K\bar{K}]}=0.42\pm0.04,
 \end{equation}
which overlaps with the experimental range (Table \ref{table1}). For $\pi \pi$ channel, decay width has the right order of magnitude compared to the full width for large values of $m[\pi(1300)]$  and
\begin{equation}
 \frac{\Gamma[f_0(1710)\rightarrow \pi \pi]}{\Gamma[f_0(1710)\rightarrow K\bar{K}]}=0.10\pm0.05,
 \end{equation}
which is not in the range of PDG data but consistent with the results of Ref. \citen{ab}:
$$\frac{\Gamma[f_0(1710)\rightarrow \pi \pi]}{\Gamma[f_0(1710)\rightarrow K\bar{K}]}<0.11.$$
Moreover the recent results of Refs. \citen{fglue1} and \citen{fglue2} have suggested that the $f_0(1710)$ is predominantly the glueball state and due to the fact that the glueball effects are not included in GLSM, we do not expect to get good results for decay widths of this state.\\ \\

\noindent{\underline{\textbf{Figure \ref{plot5}}}}:\\
\par

$ÃÂÃÂ¯ÃÂÃÂ¿ÃÂÃÂ½\Gamma[K_0^*(1430)\rightarrow \pi K]$ is too large, but according to final state interactions of $\pi K$ which are estimated by the K-matrix unitarization method, decay width decreases \cite{piK}, but this time becomes too small compared to the experiment (Table \ref{table5}) and this convinces us that the current Lagrangian of the model, cannot describe energy region above 1 GeV for some processes  and therefore is not complete.\\  \\ \\ 

\noindent{\underline{\textbf{Figure \ref{plot6}}}}:\\
\par
Among different hadronic channels for $a_0(1450)$ decay, only $\Gamma[a_0(1450) \rightarrow \pi \eta']$ is comparable with the experimental full width and the other two, i.e., $\pi \eta$ and $K\bar{K}$ widths, are too large but their ratio is in the experimental range
\begin{equation}
 \frac{\Gamma[a_0(1450)\rightarrow K \bar{K}]}{\Gamma[a_0(1450)\rightarrow \pi \eta]}=0.96\pm0.10.
 \end{equation}
 \par
It should be pointed that the final state interaction effects are also computed for $\pi \eta$ scattering \cite{pieta} and fortunately the total width for $a_0(1450)$ which comes from the imaginary part of the K-matrix unitarized amplitude pole, is close to the experimental range (Table \ref{table5}). This will be discussed in the next section.\\  \\

\noindent{\underline{\textbf{Figure \ref{plot7}}}}:\\
\par
Decay widths of different modes of $\eta(1295)$ are calculated within scenario $3I$ which is the best scenario. Decay widths for both channels are consistent with the experimental data for small values of $m[\pi(1300)]$ and their ratio is
\begin{equation}
\frac{\Gamma[\eta(1295) \rightarrow a_0(980)\pi]}{\Gamma[\eta(1295) \rightarrow\sigma \eta]}=0.47 \pm 0.16,
\end{equation}
which overlaps with the experimental range.\\ \\
\noindent{\underline{\textbf{Figure \ref{plot8}}}}:\\
\par
To evaluate $\eta(1405)$ and $ \eta(1475) \rightarrow a_0(980)\pi$ decay widths, we used scenario $5I$ and $6I$, respectively. Although $\eta$ masses predicted in these scenarios are less than expected values and these scenarios are clearly not favoured, but the resulting widths are comparable with the full experimental width (Fig. \ref{plot8}). Note that in our model, $\eta(1405)$ can not decay into $f_0(980)\eta$ as its predicted mass is less than the threshold mass.

\par
In order to compare the predictions of the model with experimental data, the averaged values and the standard deviations over the $m[\pi(1300)]-A_3/A_1$ plane for decay widths of lowest lying  scalars and pseudoscalars are given in Table \ref{table2}. Also the mean values for decay widths of  next-to-lowest lying mesons are summarized in Table \ref{table3}. Among these decays, nine of them overlap with the experimental range or at least have the right order of magnitude compared to the experimental full width. For the remaining decays, i.e., $f_0(1370)\rightarrow \pi \pi$, $f_0(1710)\rightarrow K \bar{K}$, $f_0(1710)\rightarrow \eta \eta$, $K_0^*(1430)\rightarrow \pi K $, $a_0(1450)\rightarrow \pi \eta$ and $a_0(1450)\rightarrow K \bar{K}$, decay widths are too large. 
\par
One may argue that if we consider the effect of final state interactions in  $\pi \pi$, $\pi K$ and $\pi \eta$ scatterings within the GLSM framework, this may change the bare predictions  of decay widths and  masses of scalars to the acceptable ones. In the next section, we will review this effect.
\begin{table}[!htbp]
\footnotesize
\centering
\tbl{Predicted bare masses and decay widths of \textbf{lowest lying} mesons in the generalized linear sigma model.}
{\begin{threeparttable}
{\begin{tabular}{@{}c|cccc@{}}
\noalign{\hrule height 1pt}
\noalign{\hrule height 1pt}
&   \multicolumn{2}{c}{\textbf{GLSM}} \\
             & Width (MeV) & Mass of decaying  \\
             &  &  particle (MeV) &   \\
\noalign{\hrule height 1pt}
$\sigma \rightarrow \pi \pi$ &  $531 \pm 99$ & $645 \pm 42$ \\
$f_0(980)\rightarrow \pi \pi$ &  $35 \pm 27$ & $1131 \pm 47$  \\
$f_0(980) \rightarrow  K \overline{K}$  &  $21 \pm 31$ & $1131 \pm 47$   \\
$\kappa \rightarrow \pi K$ &   $58 \pm 90$ & $1105 \pm 28$ \\
$a_0(980)\rightarrow \pi \eta $ &  $57 \pm 44$ & $980 \pm 20 \tnote{a}$   \\
$a_0(980) \rightarrow K \overline{K}$ &  $21 \pm 27$ & $980 \pm 20$   \\
\noalign{\hrule height 1pt}
\noalign{\hrule height 1pt}
\end{tabular}\label{table2}}
\begin{tablenotes}
\item[a] This shows experimental error bar;  Other error bars are standard deviations of the mean value averaged over the $m[\pi(1300)]-A_3/A_1$ plane.
\end{tablenotes}
\end{threeparttable}}
\end{table}

\begin{table}[!htbp]
\footnotesize
\centering
\tbl{Predicted bare masses and decay widths of \textbf{next-to-lowest lying} mesons in GLSM. }
{\begin{threeparttable}
{\begin{tabular}{@{}c|cc!{\vrule width 1.5pt}c|cc@{}}
\noalign{\hrule height 1pt}
\noalign{\hrule height 1pt}
\textbf{decay modes}&\multicolumn{2}{c!{\vrule width 1.5pt}}{\textbf{GLSM}}& \textbf{decay modes}&\multicolumn{2}{c}{\textbf{GLSM}} \\
            & width (MeV)  & mass of decaying& & width (MeV)  & mass of decaying  \\
             &  &  particle(MeV)& & & particle(MeV)  \\
\noalign{\hrule height 1pt}
$f_0(1370) \rightarrow \pi \pi $ &  $3216 \pm 313$& $1504 \pm 6$ & $a_0(1450) \rightarrow \pi \eta$   &   $1729 \pm 278$   &   $1474 \pm 19\tnote{a}$   \\
$f_0(1370) \rightarrow K \overline{K}$ &  $72 \pm 76$ & $1504 \pm 6$  &$a_0(1450)\rightarrow \pi \eta'$  &  $106 \pm 57$  & $1474 \pm 19$  \\
$f_0(1370) \rightarrow \eta \eta$  &  $107 \pm 98$  & $1504 \pm 6$ &$a_0(1450) \rightarrow K \overline{K}$ &  $1660 \pm 278$ & $1474 \pm 19$    \\
$f_0(1370) \rightarrow \pi(1300) \pi $ &  $59 \pm 52$ & $1504 \pm 6$ & $\eta(1295) \rightarrow a_0(980) \pi$ \tnote{b} &  $62 \pm 30$ & $1271 \pm 61$  \\
$f_0(1710) \rightarrow \pi \pi$&  $339 \pm 204$ & $1684 \pm 40$  & $\eta(1295) \rightarrow \sigma \eta$\tnote{b}  & $103\pm 72$& $1271 \pm 61$\\
$f_0(1710) \rightarrow K \overline{K}$ &  $3306 \pm 523$& $1684 \pm 40$ &  $\eta(1405) \rightarrow a_0(980) \pi$\tnote{c}  &  $65 \pm 35$ & $1274 \pm 59$  \\
$f_0(1710) \rightarrow \eta \eta$ &  $1406 \pm 337$ & $1684 \pm 40$  & $\eta(1475) \rightarrow a_0(980) \pi$ \tnote{d}& $69\pm 39$ & $1278 \pm 59$   \\
$K_0^*(1430)\rightarrow \pi K $   &  $4674 \pm 562$  & $1555 \pm 36 $  & & &  \\
\noalign{\hrule height 1pt}
\noalign{\hrule height 1pt}
\end{tabular}\label{table3}}
\begin{tablenotes}
\item[a] This shows experimental error bar$ ; $ Other error bars are standard deviations of the mean value averaged over the $m[\pi(1300)]-A_3/A_1$ plane.
\item[b] Scenario $ 3I $
\item[c] Scenario $ 5I $
\item[d] Scenario $ 6I $
\end{tablenotes}
\end{threeparttable}}
\end{table}


\section{Unitarity Corrections}\label{sec4}

In order to consider final state interactions in $\pi \pi$, $\pi K$ and $\pi \eta$ scatterings, we use K-matrix unitarization method \cite{pipi,piK,pieta} through which the partial wave bare amplitude $T^{I\, B}_{l}$ transforms to unitarized amplitude $T^I_l$ using the following equation
\begin{equation}
T^I_l=\frac{T^{I\,B}_{l}}{1-i T^{I\, B}_{l}},
\end{equation}
where $I$ and $l$ denote to the partial wave isospin and angular momentum. The poles of the K-matrix unitarized
amplitude are used to calculate the physical masses and full decay widths of the intermediate scalar mesons. 
Here we review the $\pi\pi$ scattering \cite{pipi} and the $\pi K$ and $\pi \eta$ scatterings are explored in Refs. \citen{piK,pieta}.
\par
\begin{figure}[!htbp]
\begin{center}
\vskip 1cm
\epsfxsize = 7.5cm
 \includegraphics[height=6 cm]{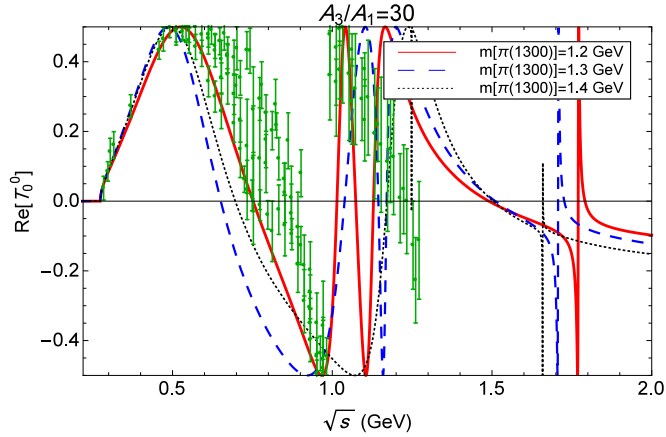}
 \caption{ Real part of the unitarized scattering amplitude of $\pi\pi$ scattering for three different choices of $m[\pi(1300)]$ and $A_3/A_1=30$.}
 \label{pipiscat}
\end{center}
\end{figure}
\begin{figure}[!htbp]
\begin{center}
\vskip 0.5 cm
\epsfxsize = 5 cm
 \includegraphics[height=6 cm]{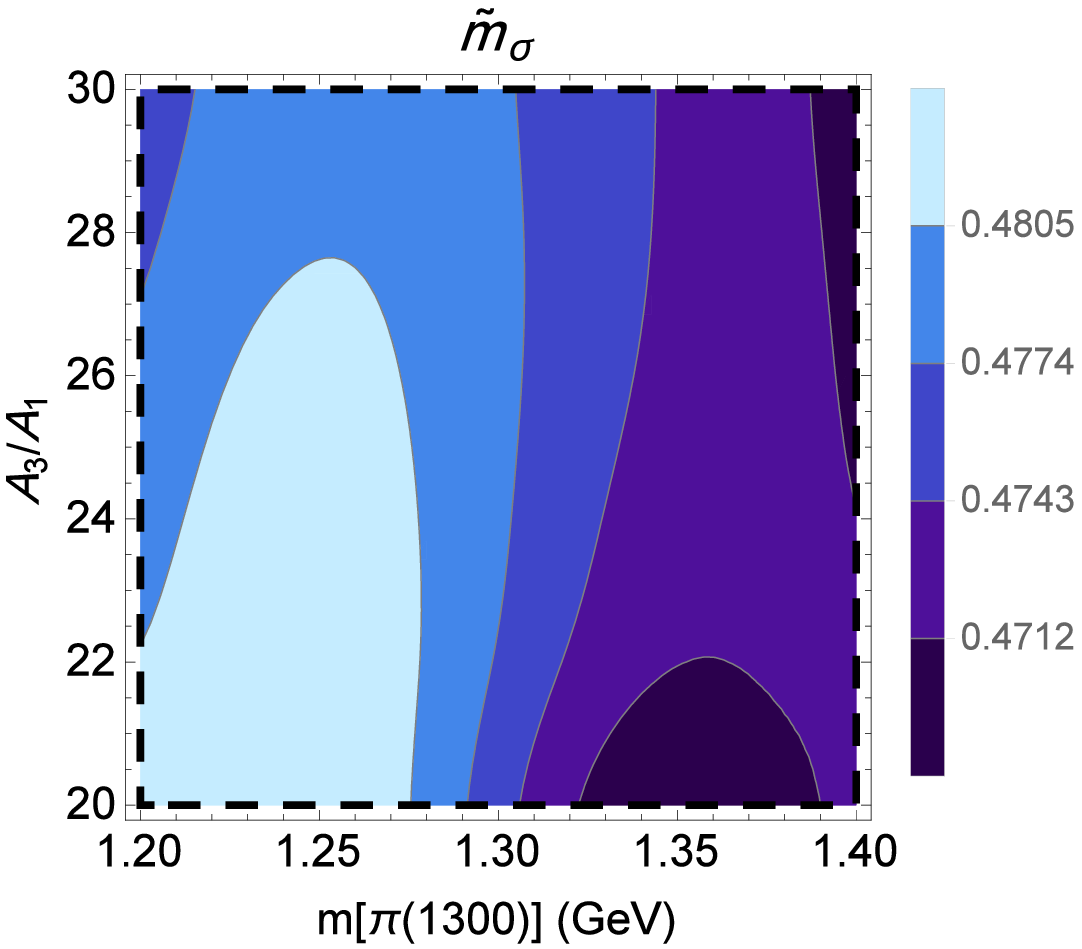}
\hskip 1cm
\epsfxsize = 5 cm
 \includegraphics[height=6 cm]{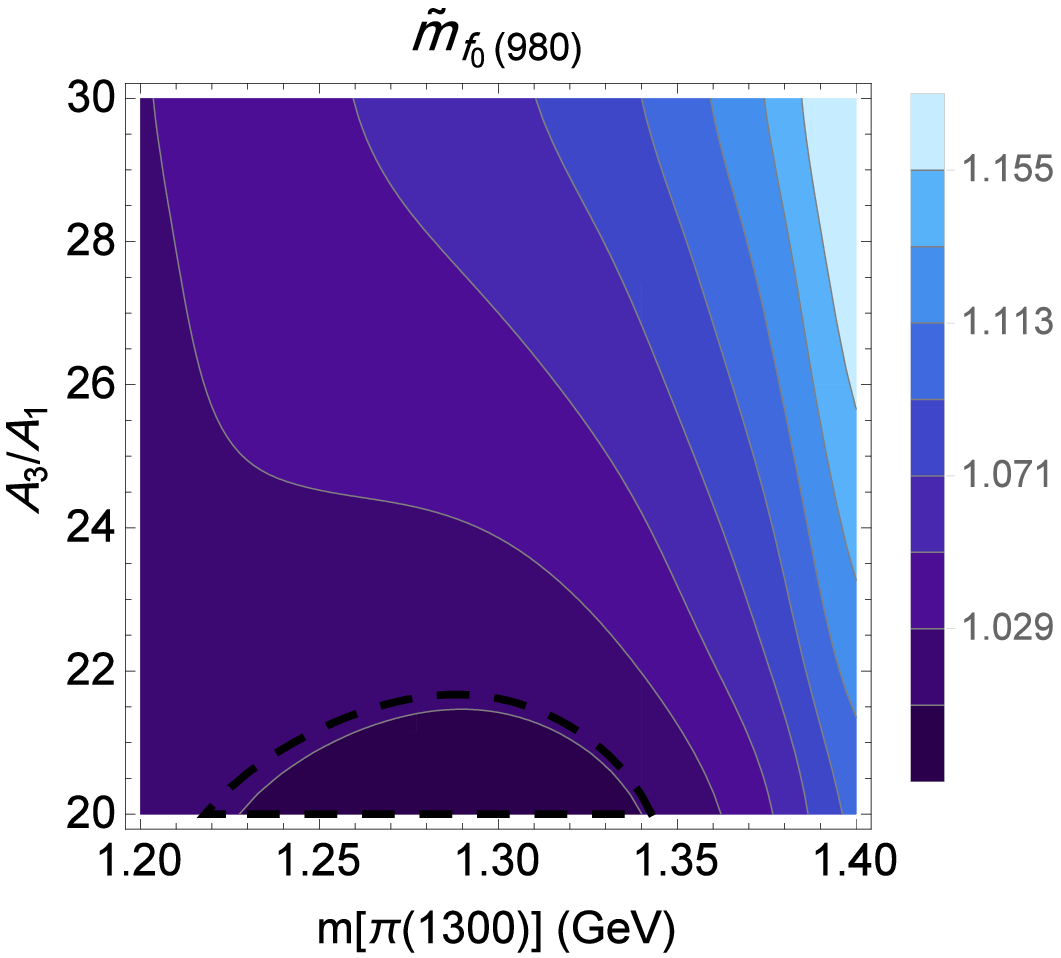}
 \vskip 0.5 cm
\epsfxsize = 5 cm
 \includegraphics[height=6 cm]{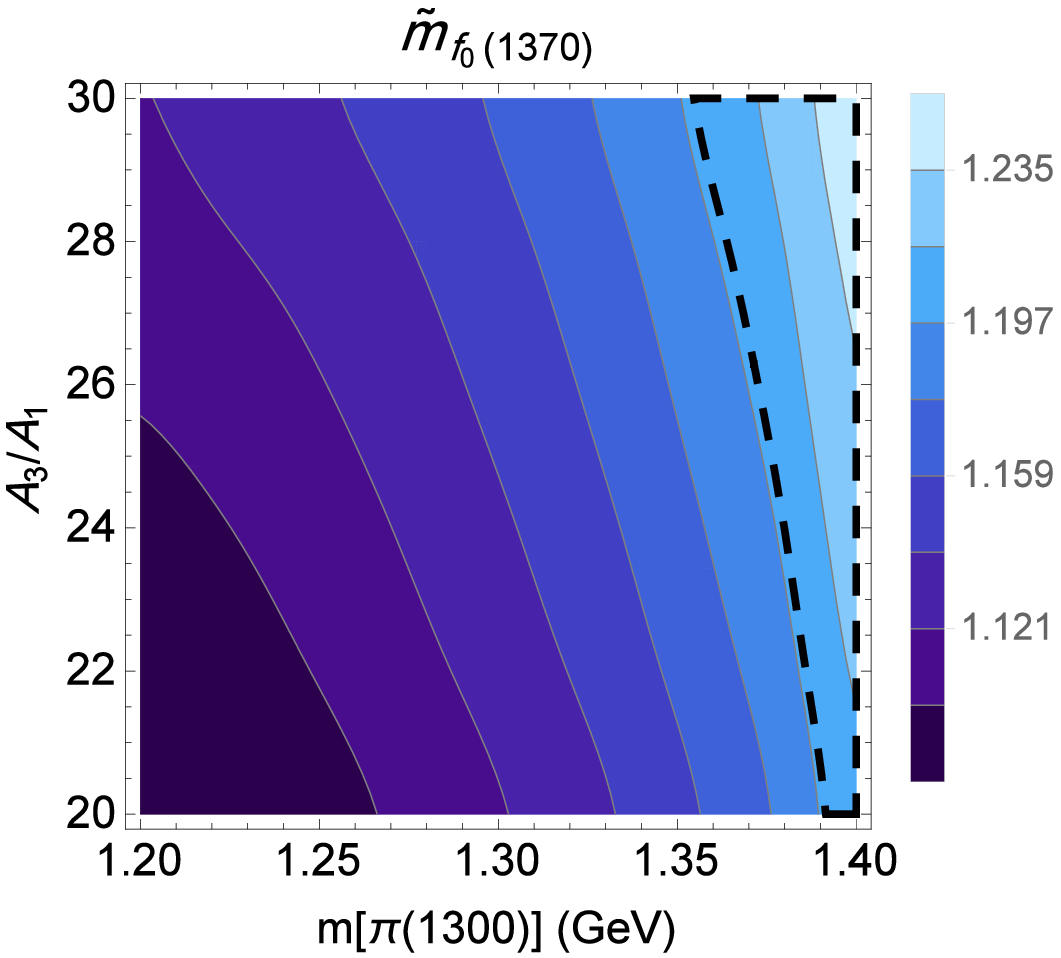}
  \hskip 1cm
\epsfxsize = 5 cm
 \includegraphics[height=6 cm]{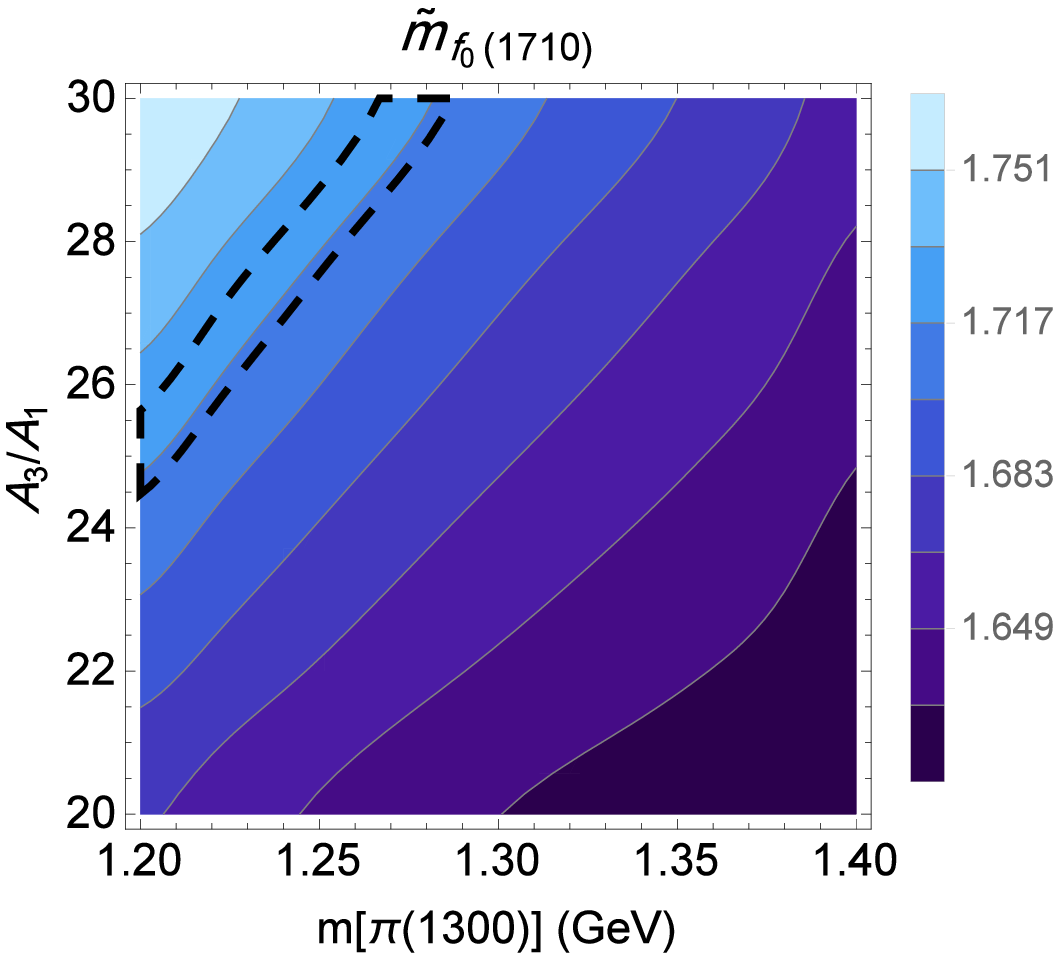}
 \caption{Contour plots of the predictions of the GLSM for the physical masses of four f-mesons obtained from the poles of the K-matrix unitarized amplitude of $\pi\pi$ scattering over the $m[\pi(1300)]-A_3/A_1$ plane. The parameter space inside the dashed curves indicate regions for which masses are in the experimental range. }
 \label{pipi_pole_mass}
\end{center}
\end{figure}
\begin{figure}[!htbp]
\begin{center}
\vskip 1cm
\epsfxsize = 7.5cm
 \includegraphics[height=6 cm]{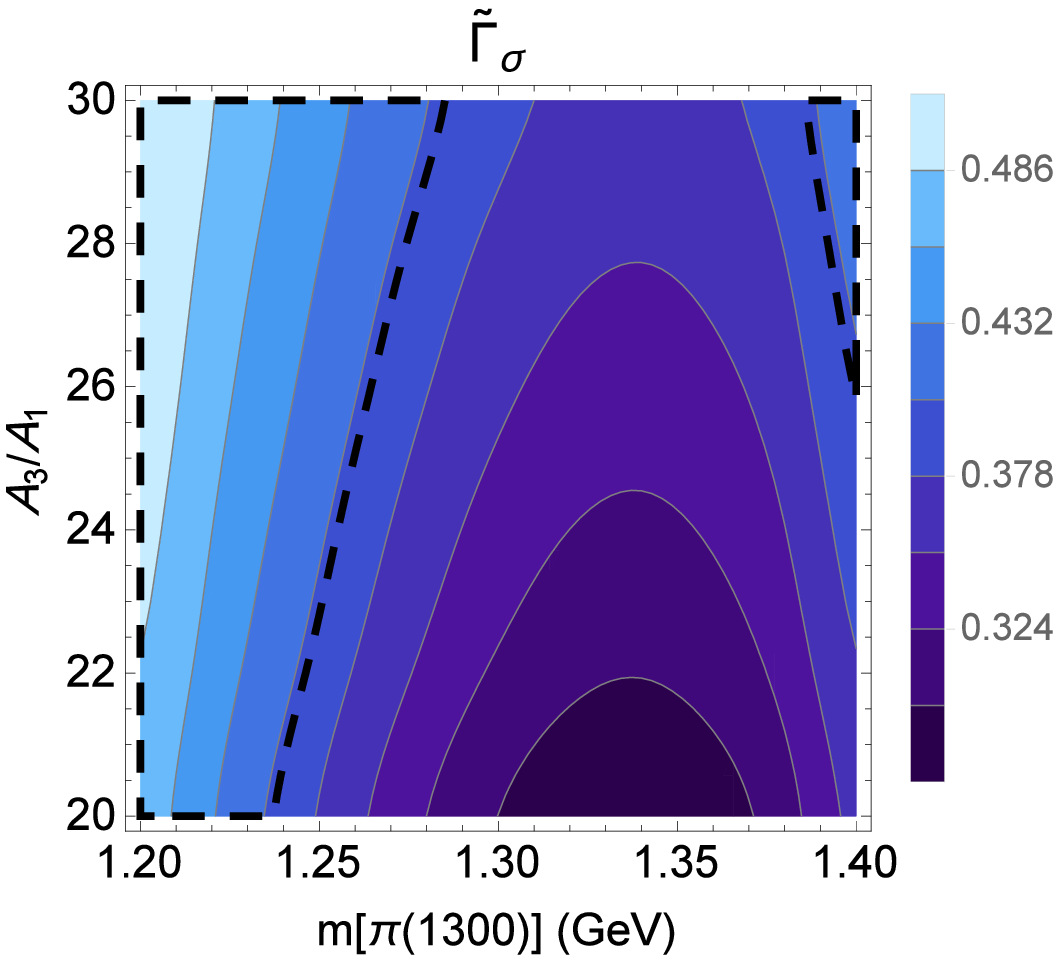}
\hskip 1cm
\epsfxsize = 7.5cm
 \includegraphics[height=6 cm]{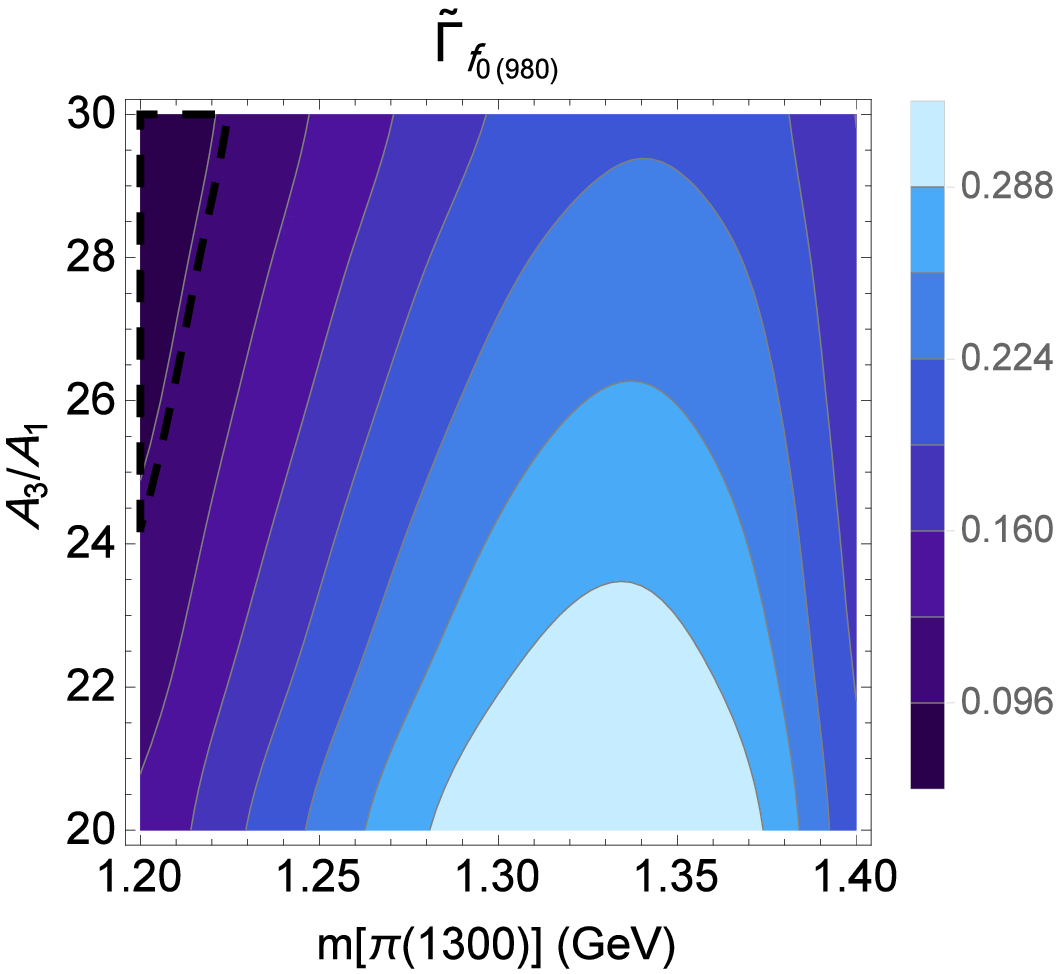}
 \vskip 1cm
\epsfxsize = 7.5cm
 \includegraphics[height=6 cm]{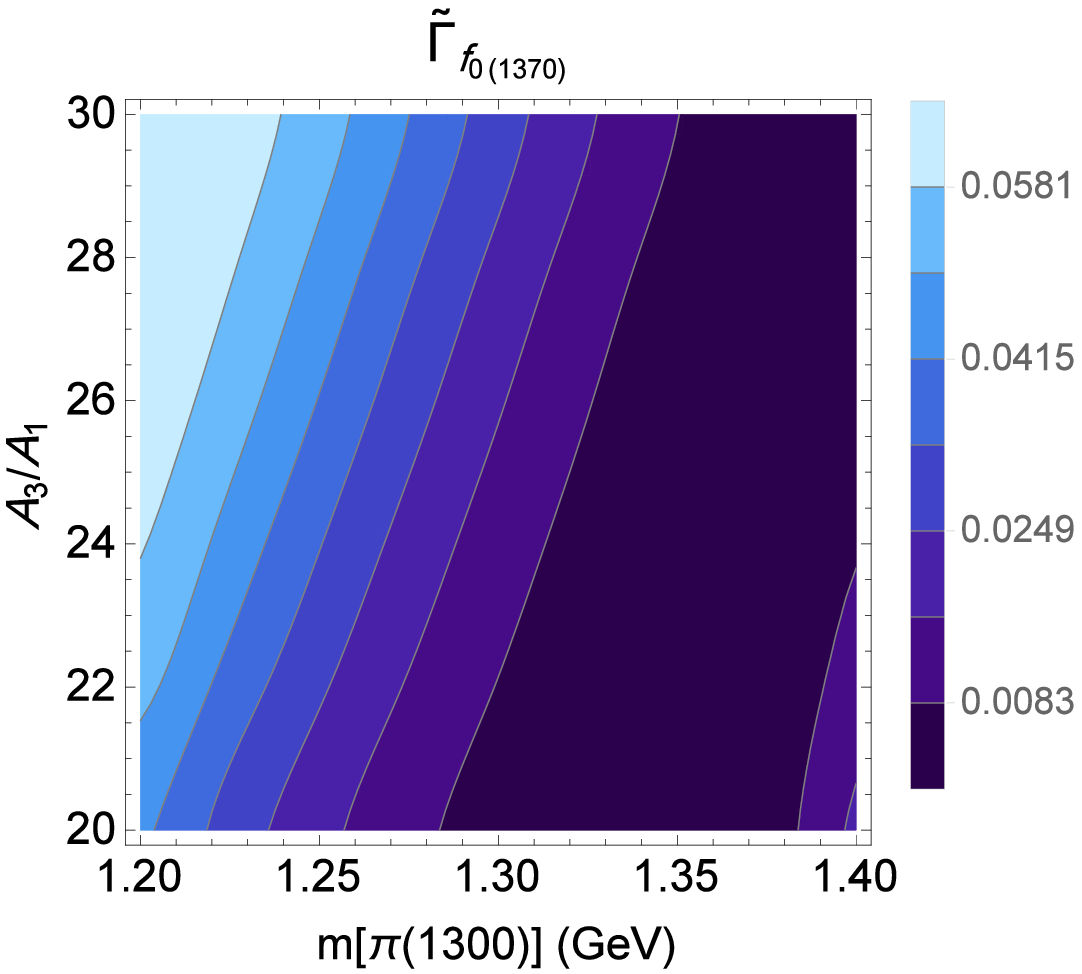}
  \hskip 1cm
\epsfxsize = 7.5cm
 \includegraphics[height=6 cm]{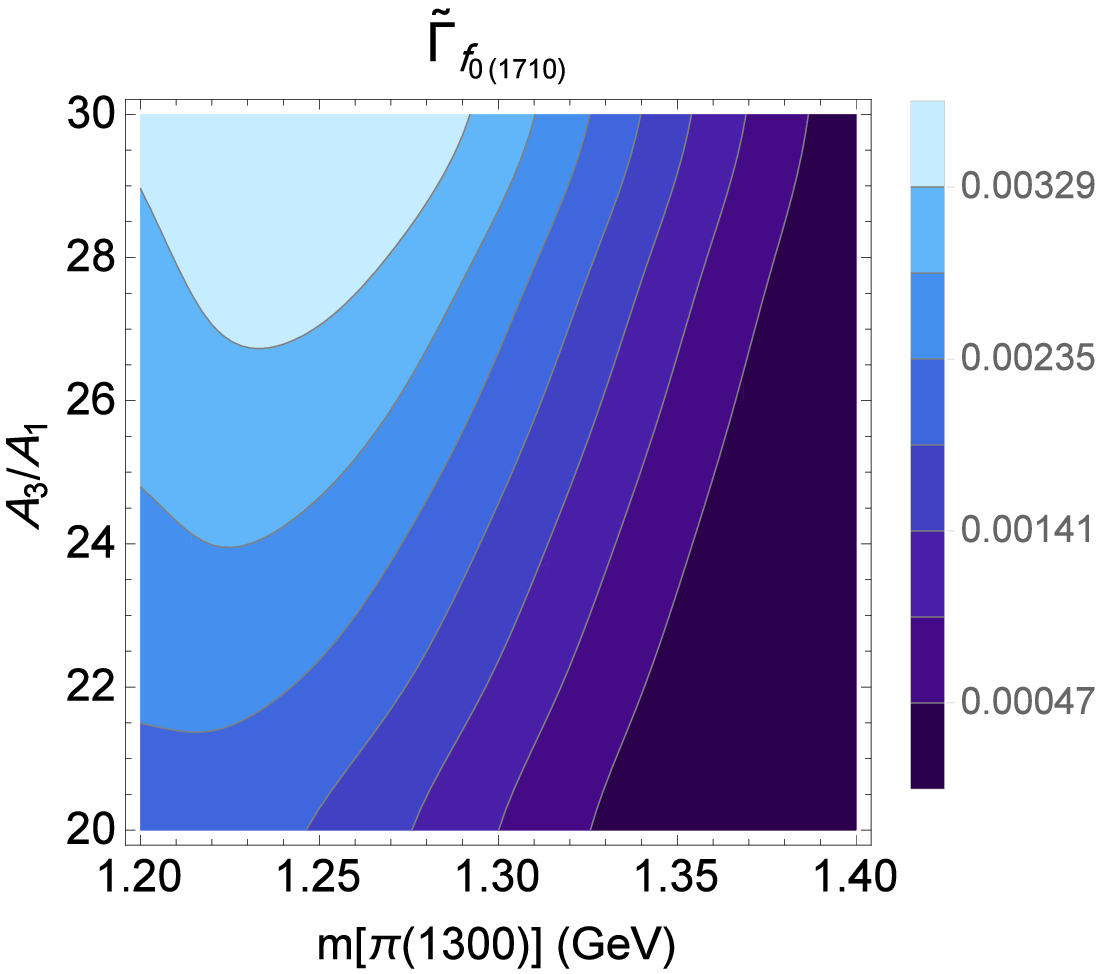}
 \caption{Contour plots of the predictions of the GLSM for the physical widths of four f-mesons obtained from the poles of the K-matrix unitarized amplitude of $\pi\pi$ scattering over the $m[\pi(1300)]-A_3/A_1$ plane. The parameter space inside the dashed curves indicate regions for which decay widths are in the experimental range. }
 \label{pipi_pole_width}
\end{center}
\end{figure}
\begin{figure}[!htbp]
\begin{center}
\vskip 1cm
\epsfxsize = 7.5cm
 \includegraphics[height=6 cm]{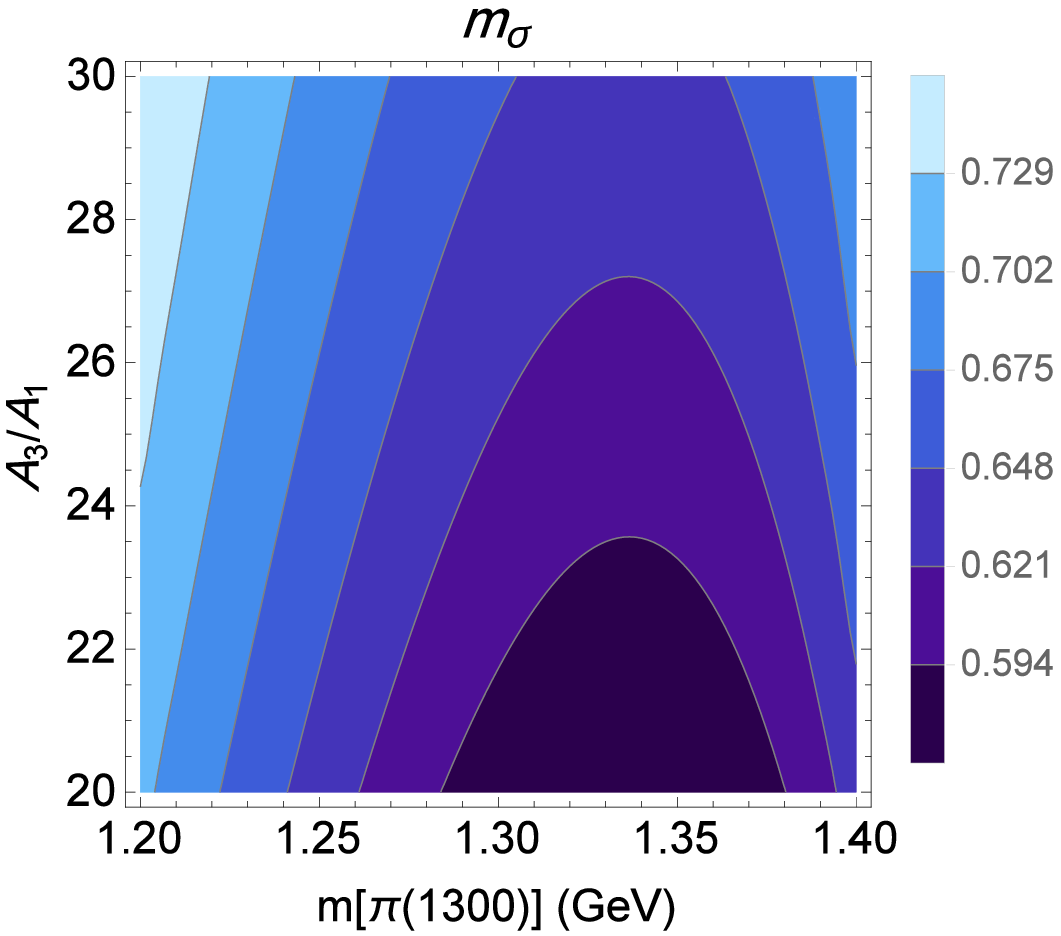}
\hskip 1cm
\epsfxsize = 7.5cm
 \includegraphics[height=6 cm]{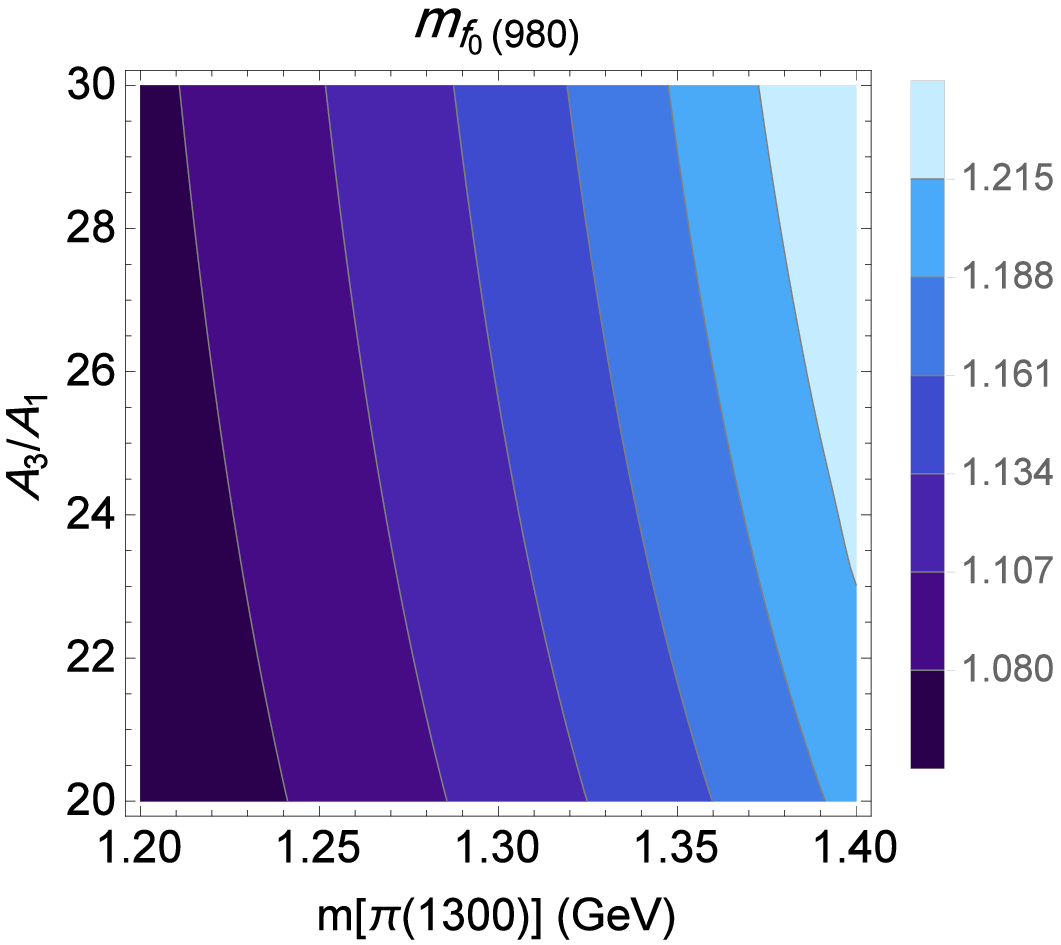}
 \vskip 1cm
\epsfxsize = 7.5cm
 \includegraphics[height=6 cm]{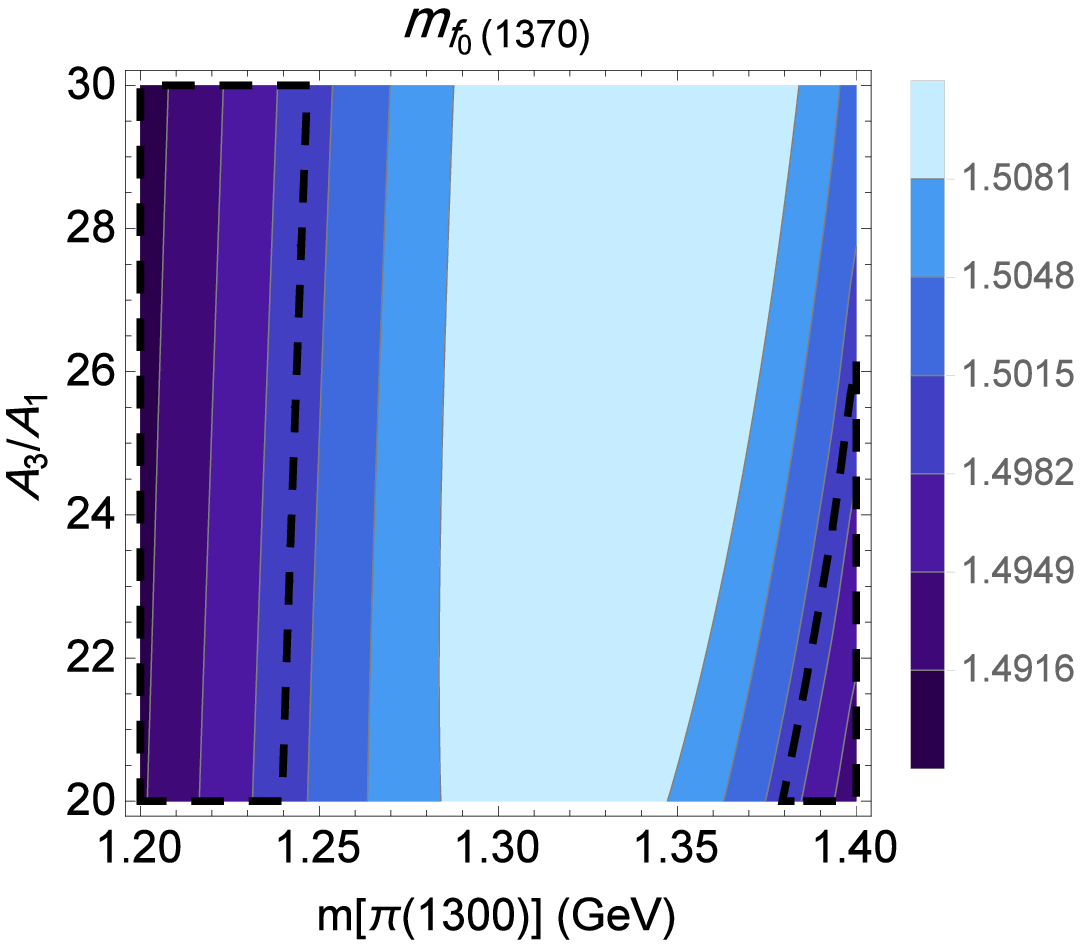}
  \hskip 1cm
\epsfxsize = 7.5cm
 \includegraphics[height=6 cm]{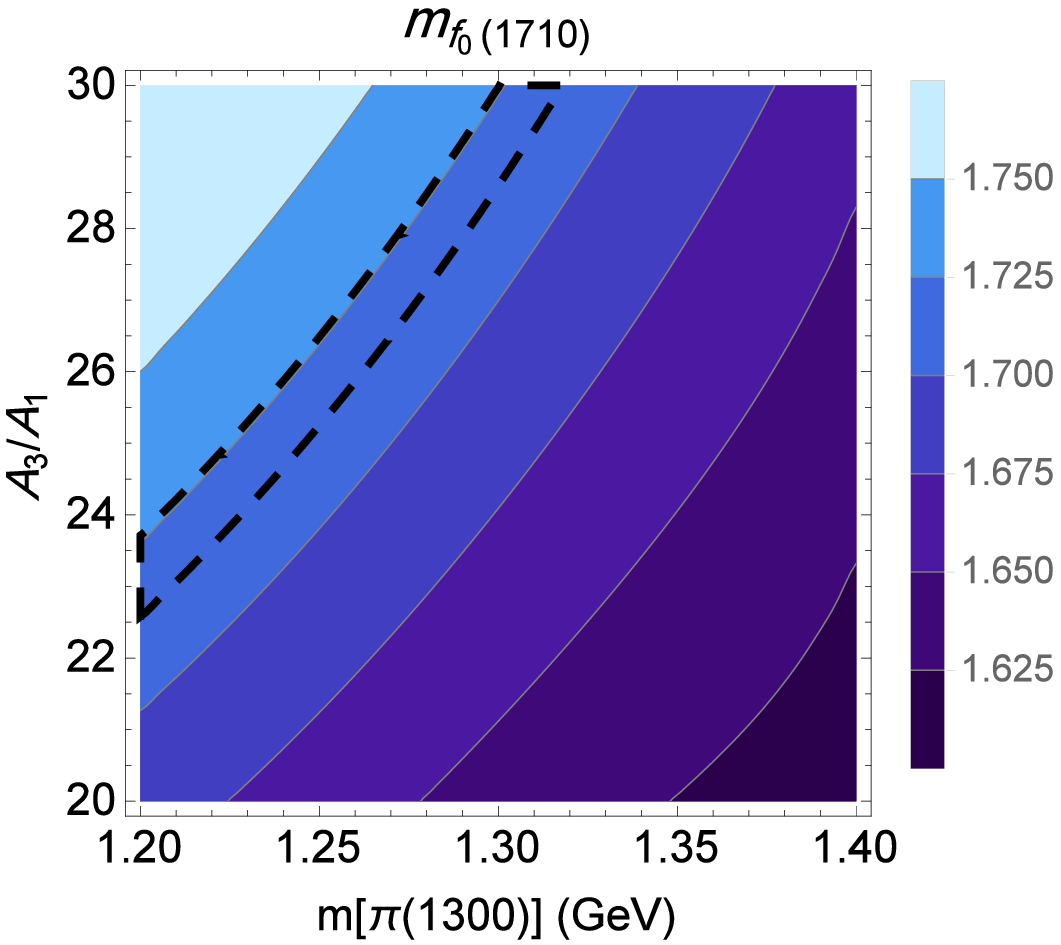}
 \caption{Contour plots of the predictions of the GLSM for the bare masses of four f-mesons over the $m[\pi(1300)]-A_3/A_1$ plane. The parameter space inside the dashed curves indicate regions for which decay widths are in the experimental range. }
 \label{f_bare_mass}
\end{center}
\end{figure}

The K-matrix unitarized $\pi\pi$ scattering amplitude  for the $I=J=0$ channel  is given by
\begin{equation}
T_0^0 = {  {T_0^0}^B \over { 1 - i\, {T_0^0}^B} },
\label{T00_unitary}
\end{equation}
where ${T_0^0}^B$ is the ``bare'' scattering amplitude calculated from the Lagrangian in Eq. (\ref{mixingLsMLag})
\begin{equation}
{T_0^0}^B = T_\alpha + \sum_i  {  {T_\beta^i } \over
{m_{f_i}^2 - s}},
\label{T00B}
\end{equation}
with
\begin{eqnarray}
T_\alpha &=&
{1\over 64 \pi}
\sqrt{1 - {4 m_\pi^2\over s}}\,
\left[-5\, \gamma^{(4)}_{\pi\pi} +
  { 2 \over {p_\pi^2}}\,
   \sum_i \gamma_{f_i\pi\pi}^2\,  {\rm ln} \left(1 +  {{4
p_\pi^2}\over m_{f_i}^2} \right)
\right],
\nonumber \\
T_\beta^i &=&
{3\over 16 \pi}
\sqrt{1 - {4 m_\pi^2\over s}}\, \gamma_{f_i\pi\pi}^2,
\label{pipiformula}
\end{eqnarray}
where $p_\pi = \sqrt{s - 4 m_\pi^2} / 2$. The scalar-pseudoscalar-pseudoscalar couplings $\gamma_{f_i\pi\pi}$ are defined in Sec. \ref{sec3}, and $\gamma^{(4)}_{\pi\pi}$ is  the pion four-point coupling constant
\begin{equation}
g
 =
\left\langle
{{\partial^4 V}
\over
{\partial \pi^+ \, \partial \pi^- \, \partial \pi^+ \,
\partial \pi^-}}
\right\rangle
 =
\sum_{A,B,C,D}
\left\langle
{{\partial^4 V}
\over
{
 \partial (\phi_1^2)_A \,
 \partial (\phi_2^1)_B \,
 \partial (\phi_1^2)_C \,
 \partial (\phi_2^1)_D
}}
\right\rangle \,
(R_\pi)_{A1} \,
(R_\pi)_{B1} \,
(R_\pi)_{C1} \,
(R_\pi)_{D1},
\end{equation}
where the sum is over ``bare'' pions and
$A, B, \cdots$ = 1, 2 with 1 denoting nonet $M$ and 2
denoting nonet $M'$ and  $R_\pi$ is the pion
rotation matrix. The real part of the unitarized scattering amplitude, Eq. (\ref{T00_unitary}), is plotted in Fig \ref{pipiscat}. It is clear that the prediction of the model  is in reasonable qualitative agreement with the experimental data up to about 1 GeV.
The physical pole positions in the unitarized scattering amplitude are determined by solving for the complex roots of the denominator of the K-matrix unitarized amplitude Eq. (\ref{T00_unitary})
\begin{equation}\label{eqpole}
1- i T_0^{0B}=0.
\end{equation}
Each complex pole is equivalent to ${\tilde m}_i^2 - i {\tilde m} {\tilde \Gamma}_i$, where ${\tilde m}_i$ and ${\tilde \Gamma}_i$ are the physical mass and width of the $i$-th pole. By solving Eq. (\ref{eqpole}) numerically, we will find four solutions for the pole positions, each corresponds to the physical mass and total width of $f_i$'s. The masses and total widths of the $\pi\pi$ unitarized scattering amplitude are plotted in Figs. \ref{pipi_pole_mass} and \ref{pipi_pole_width}. Comparing the bare masses resulted from the Lagrangian (Fig. \ref{f_bare_mass}) with the physical masses obtained from the unitarized $\pi\pi$ scattering amplitude (Fig. \ref{pipi_pole_mass}), shows that the unitarization procedure reduces the bare masses of $\sigma$ and $f_0(980)$ which are located out of the experimental ranges to physical masses within the experimental bounds. It is clear from Figs.  \ref{plot1} and \ref{pipi_pole_width} that before and after unitarization there  exists regions of parameter space for which decay widths of $\sigma$ and $f_0(980)$ cover the experimental ranges \footnote{As $\pi\pi$ channel is the dominant decay mode for $\sigma$ and $f_0(980)$ decays, total widths resulted from unitarization procedure are comparable with $\Gamma[\sigma \rightarrow \pi \pi]$ and $\Gamma[f_0(980) \rightarrow \pi \pi]$.}.  Furthermore, for $f_0(1370)$ and $f_0(1710)$, the bare total widths (sum of the average values for different decay modes in Table \ref{table3}), are too large compared to the experimental full width which proves the failure of the model for these decay widths.  As we expected, the unitarization, makes the total widths smaller with respect to the previous values, but unfortunately too small compared to the experimental data (Fig. \ref{pipi_pole_width}). The same unitarization procedure can be applied for $\pi K$ and $\pi \eta$ scatterings which leads to the unitarized masses and full widths for isodoublets ($\kappa$, $K_0^*$) and isotriplets ($a_0(980)$,$a_0(1450)$), respectively \cite{piK,pieta}.

\par
Finally, the mean values for physical masses and decay widths obtained from unitarization procedure corresponding to isosinglet, isodoublet and isotriplet scalar states  are given in Table \ref{table5}.  It is clear that the averaged total widths of $\sigma$ and $a_0(980)$  with their uncertainty overlap with the experiment. For $f_0(980)$, although  the mean value of $\tilde{\Gamma}_{f_0(980)}$  is not in agreement with the experimental data, it varies from $75$ to $335$ MeV which  indeed for some  regions of parameter space, total width overlaps with PDG data.  $\tilde{\Gamma}_{\kappa}$ does not agree with the experiment but is close to it. Thus, the poles in the unitarized scattering amplitude which represent physical states leads to the acceptable widths and masses for states below 1 GeV, while for the physical states above 1 GeV, the obtained total widths are too small compared to the expected values except for $a_0(1450)$.

\begin{table}[!htbp]
\small
\centering
\tbl{Predicted physical masses and decay widths of lowest lying and next-to-lowest lying mesons in GLSM obtained from unitarized amplitudes \cite{pipi,piK,pieta}.}
{\begin{tabular}{@{}c|cc!{\vrule width 1.5pt}c|cc@{}}
\noalign{\hrule height 1pt}
\noalign{\hrule height 1pt}
 & \multicolumn{2}{c!{\vrule width 1.5pt}}{\textbf{lowest lying}} & & \multicolumn{2}{c}{\textbf{next-to-lowest lying}}\\
& Width (MeV)  & Mass(MeV) &  & Width (MeV)  & Mass(MeV)  \\
  \noalign{\hrule height 1pt}
$\sigma $   &  $385 \pm 61$ & $476 \pm 4$   & $f_0(1370)$               &  $22 \pm 21$       & $1149 \pm 43$   \\
$f_0(980)$  &    $207 \pm 65$ & $1053 \pm 44$  & $f_0(1710)$               &  $2 \pm 1$           & $1672 \pm 37$    \\
$\kappa $   &    $689 \pm 27$ & $722 \pm 28$  & $K_0^* (1430) $       &  $5 \pm 6$             & $1114 \pm 50$    \\
$a_0(980) $ &    $60 \pm 52$  & $984 \pm 7$ & $a_0(1450) $             &  $502 \pm 88$       & $1085 \pm 43$   \\
\noalign{\hrule height 1pt}
\noalign{\hrule height 1pt}
\end{tabular}\label{table5}}
\end{table}

\section{Summary and conclusions}\label{sec5}

\begin{table}[!htbp]
\footnotesize
\centering
\tbl{Predicted bare masses and decay widths of \textbf{lowest lying} mesons in the non-renormalizable SU(3) single nonet linear sigma
model (first column), the generalized linear sigma model (second column) and the decoupling limit of GLSM (third column). }
{\begin{threeparttable}
{\begin{tabular}{@{}c|cc|cc|cc@{}}
\noalign{\hrule height 1pt}
\noalign{\hrule height 1pt}
& \multicolumn{2}{c|}{\textbf{SNLSM}} & \multicolumn{2}{c|}{\textbf{GLSM}} & \multicolumn{2}{c}{\textbf{GLSM in decoupling limit}}\\
            & Width (MeV)  & Mass of decaying & Width (MeV) & Mass of decaying & Width (MeV) & Mass of decaying  \\
             &  &  particle (MeV) &  &  particle (MeV) &  &  particle (MeV)  \\
\noalign{\hrule height 1pt}
$\sigma \rightarrow \pi \pi$ &  $830$ & $847$ &  $531 \pm 99$ & $645 \pm 42$ &  $3077 \tnote{a} $ & $960 \tnote{a}  $ \\
$f_0(980)\rightarrow \pi \pi$ &  $4109$ & $1300$ &  $35 \pm 27$ & $1131 \pm 47$&  $...\tnote{b}$ & $1975 \pm 117$  \\
$f_0(980) \rightarrow  K \overline{K}$ &  $1525$ & $1300$ &  $21 \pm 31$ & $1131 \pm 47$ &  $2424 \pm 46$ & $1975 \pm 117$   \\
$\kappa \rightarrow \pi K$ &  $2350$ & $1300$ &  $58 \pm 90$ & $1105 \pm 28$&  $1844 \pm 30$ & $1202 \pm 22$  \\
$a_0(980)\rightarrow \pi \eta $ &  $381$ & $1100$ &  $57 \pm 44$ & $980 \pm 20 \tnote{c}$  &  $347 \pm 16$ & $980 \pm 20 $  \\
$a_0(980) \rightarrow K \overline{K}$ &  $221$ & $1100$ &  $21 \pm 27$ & $980 \pm 20$  &  $161 \pm 100$ & $980 \pm 20$  \\
\noalign{\hrule height 1pt}
\noalign{\hrule height 1pt}
\end{tabular}\label{table4}}
\begin{tablenotes}
\item[a] These values are independent of $A_3/A_1$ and therefore have no standard deviations.
\item[b] In the decoupling limit $f_0(980)$ contains strange quarks and thus cannot couple with $\pi$, i.e., $\gamma_{f_2\pi\pi}=0$.
\item[c] This shows experimental error bar;  Other error bars are standard deviations of the mean value averaged over the $m[\pi(1300)]-A_3/A_1$ plane in case of GLSM or averaged over different values of $A_3/A_1$ in case of GLSM in decoupling limit.
\end{tablenotes}
\end{threeparttable}}
\end{table}

\noindent
In the present work, we have studied  the effect of underlying mixing between $q\bar{q}$ and $qq\bar{q}\bar{q}$ components in hadronic two-body decays of scalar and pseudoscalar mesons below 2 GeV within the GLSM. To see whether the mixing effect could improve the prediction of SNLSM for lowest lying scalars \cite{unitarize}, a comparison is made between predicted decay widths evaluated in these models ( first two columns of Table \ref{table4} ). At first sight, it is clear that the predictions have greatly improved in GLSM.   So this model  not only determines  quark structure of mesons but also can improve the predictions for decays and certainly scatterings of mesons \cite{pipi,piK,pieta}. \\
We should note that the masses of $\sigma$, $f_0(980)$ and their mixing angle $\theta_s$,  the mass of $a_0(980)$ in the isotopic spin invariant limit and therefore the related vertices such as $\gamma_{\sigma \pi \pi}$ and  $\gamma_{f_2 \pi \pi}$ are not predicted in SNLSM directly: The isoscalar masses ($m[\sigma]$, $m[f_0(980)]$) and consequently widths in Table \ref{table4}  comes from the best fit for the real part of the $I=J=0$, $\pi \pi$ scattering amplitude. Also the isovector mass ($m[a_0(980)$]) can not be predicted from the SNLSM and an arbitrary value should be chosen for it; But in GLSM, all the masses and widths are direct predictions of the model and this is another advantage of GLSM versus SNLSM.

\par
We have also evaluated all the lowest lying decay widths in the decoupling limit \footnote{Further discussion on decoupling limit is given in  \ref{deli}} ($d_2,\, e_3^a \rightarrow 0$ and $\gamma_1 \rightarrow 1$) given in the third column of Table \ref{table4} in order to show 
 that mixing provides much better description of decay widths.   It is seen that by decoupling the four-quark fields,  the results go far from the experimental data.

\par
As a result, it has become clear that the GLSM Lagrangian at the present order of $N=8$ with $8$ experimental inputs and without  any further tuning is successful in predicting decay widths of lowest lying mesons, while the predictions of SNLSM model are far from the expected values. Therefore, inclusion of the underlying mixings considerably improves the results for the decay widths.
\par
We should notice that GLSM can not give acceptable widths for some states above 1 GeV. It would be worthwhile to study the effects of adding terms with higher than eight quark and antiquark lines and also considering the effects of scalar and pseudoscalar glueballs in potential. Our initial estimate shows that adding scalar and pseudoscalar glueballs has more considerable effects than adding terms with higher order of $N$.

\section*{Acknowledgments}
Authors would like to express sincere thanks towards A. H. Fariborz without whom it was impossible to accomplish this research.
\newpage
\appendix
\section{Three-point bare couplings}

\begin{eqnarray}
\left\langle\frac{\partial^{3}V}{\partial (S_{1}^{2})_{1}\partial(\phi_{2}^{1})_{1}\partial \eta_{a}}\right\rangle &=&
\frac{4 \sqrt{2}\Big (2 c_4^a \alpha _1^5 \beta _1+c_4^a \alpha _1^4 \alpha _3 \beta _3
+2 c_3 \alpha _3 \beta _3 \gamma
_1^2+2 c_3 \alpha _1 \beta _1 \gamma _1 (1+\gamma _1)\Big)}{\alpha _1^3 (2 \alpha _1 \beta _1+\alpha _3 \beta_3)},\\ \nonumber\\
\left\langle\frac{\partial^{3}V}{\partial (S_{1}^{2})_{1}\partial(\phi_{2}^{1})_{2}\partial \eta_{a}}\right\rangle&=&
-\frac{8 \sqrt{2} c_3 \left(-1+\gamma _1\right) \Big (\alpha _3 \beta _3 \gamma _1+\alpha _1 \beta _1 \left(1+\gamma _1\right)\Big )}{\alpha
_1 \left(2 \alpha _1 \beta _1+\alpha _3 \beta _3\right){}^2},\\ \nonumber\\
\left\langle\frac{\partial^{3}V}{\partial (S_{1}^{2})_{2}\partial(\phi_{2}^{1})_{1}\partial \eta_{a}}\right\rangle&=&
\frac{8 \sqrt{2} c_3\left(-1+\gamma _1\right) \Big (\alpha _3 \beta _3 \gamma _1+\alpha _1 \beta _1 \left(1+\gamma _1\right)\Big )}{\alpha
_1 \left(2 \alpha _1 \beta _1+\alpha _3 \beta _3\right){}^2},\\ \nonumber\\
\left\langle\frac{\partial^{3}V}{\partial (S_{1}^{2})_{1}\partial(\phi_{2}^{1})_{1}\partial \eta_{b}}\right\rangle&=&
\frac{8 c_3 \gamma _1 \left(\alpha _3 \beta _3+2 \alpha _1 \beta _1 \gamma _1\right)}{\alpha _1^2 \alpha _3 \left(2 \alpha _1
\beta _1+\alpha _3 \beta _3\right)},\\ \nonumber\\
\left\langle\frac{\partial^{3}V}{\partial (S_{1}^{2})_{1}\partial(\phi_{2}^{1})_{2}\partial \eta_{b}}\right\rangle&=&
4 e_3^a-\frac{8 c_3 \left(-1+\gamma _1\right) \left(\alpha _3 \beta _3+2 \alpha _1 \beta _1 \gamma _1\right)}{\alpha _3 \left(2
\alpha _1 \beta _1+\alpha _3 \beta _3\right){}^2},\\ \nonumber\\
\left\langle\frac{\partial^{3}V}{\partial (S_{1}^{2})_{2}\partial(\phi_{2}^{1})_{1}\partial \eta_{b}}\right\rangle&=&
4 e_3^a+\frac{8 c_3 \left(-1+\gamma _1\right) \left(\alpha _3 \beta _3+2 \alpha _1 \beta _1 \gamma _1\right)}{\alpha _3 \left(2
\alpha _1 \beta _1+\alpha _3 \beta _3\right){}^2},\\ \nonumber\\
\left\langle\frac{\partial^{3}V}{\partial (S_{1}^{2})_{1}\partial(\phi_{2}^{1})_{1}\partial \eta_{c}}\right\rangle&=&
\frac{8 \sqrt{2} c_3 \left(-1+\gamma _1\right) \gamma _1}{\alpha _1 \left(2 \alpha _1 \beta _1+\alpha _3 \beta _3\right)},\\ \nonumber\\
\left\langle\frac{\partial^{3}V}{\partial (S_{1}^{2})_{1}\partial(\phi_{2}^{1})_{2}\partial \eta_{c}}\right\rangle&=&
-\frac{8 \sqrt{2} c_3 \alpha _1 \left(-1+\gamma _1\right){}^2}{\left(2 \alpha _1 \beta _1+\alpha _3 \beta _3\right){}^2},\\ \nonumber\\
\left\langle\frac{\partial^{3}V}{\partial (S_{1}^{2})_{2}\partial(\phi_{2}^{1})_{1}\partial \eta_{c}}\right\rangle&=&
\frac{8 \sqrt{2} c_3 \alpha _1 \left(-1+\gamma _1\right){}^2}{\left(2 \alpha _1 \beta _1+\alpha _3 \beta _3\right){}^2},\\ \nonumber\\
\left\langle\frac{\partial^{3}V}{\partial (S_{1}^{2})_{1}\partial(\phi_{2}^{1})_{1}\partial \eta_{d}}\right\rangle&=&
\frac{8 e_3^a \alpha _1^3 \beta _1+4 e_3^a \alpha _1^2 \alpha _3 \beta _3+8 c_3 \alpha _3 \left(-1+\gamma _1\right)
\gamma _1}{\alpha _1^2 \left(2 \alpha _1 \beta _1+\alpha _3 \beta _3\right)},\\ \nonumber\\
\left\langle\frac{\partial^{3}V}{\partial (S_{1}^{2})_{1}\partial(\phi_{2}^{1})_{2}\partial \eta_{d}}\right\rangle&=&
-\frac{8 c_3 \alpha _3 \left(-1+\gamma _1\right){}^2}{\left(2 \alpha _1 \beta _1+\alpha _3 \beta _3\right){}^2},\\ \nonumber\\
\left\langle\frac{\partial^{3}V}{\partial (S_{1}^{2})_{2}\partial(\phi_{2}^{1})_{1}\partial \eta_{d}}\right\rangle&=&
\frac{8 c_3 \alpha _3 \left(-1+\gamma _1\right){}^2}{\left(2 \alpha _1 \beta _1+\alpha _3 \beta _3\right){}^2},\\ \nonumber\\
\left\langle\frac{\partial^{3}V}{\partial (S_{2}^{3})_{1}\partial(\phi_{1}^{2})_{1}\partial(\phi_{3}^{1})_{1}}\right\rangle&=&
4 \alpha_{3} c_4^a,\\ \nonumber\\
\left\langle\frac{\partial^{3}V}{\partial (S_{2}^{3})_{2}\partial(\phi_{1}^{2})_{1}\partial(\phi_{3}^{1})_{1}}\right\rangle&=&
\left\langle\frac{\partial^{3}V}{\partial (S_{2}^{3})_{1}\partial(\phi_{1}^{2})_{1}\partial(\phi_{3}^{1})_{2}}\right\rangle=
\left\langle\frac{\partial^{3}V}{\partial (S_{2}^{3})_{1}\partial(\phi_{1}^{2})_{2}\partial(\phi_{3}^{1})_{1}}\right\rangle=
-4 e_3^a,\\ \nonumber\\
\left\langle\frac{\partial^{3}V}{\partial f_a\partial(\phi_{1}^{3})_{1}\partial(\phi_{3}^{1})_{1}}\right\rangle&=&
2 \sqrt{2} c_4^a(2\alpha_1-\alpha_3),\\ \nonumber\\
\left\langle\frac{\partial^{3}V}{\partial f_b\partial(\phi_{1}^{3})_{1}\partial(\phi_{3}^{1})_{1}}\right\rangle&=&
-4 c_4^a(\alpha_1-2\alpha_3),\\ \nonumber\\
\left\langle\frac{\partial^{3}V}{\partial f_c\partial(\phi_{1}^{3})_{1}\partial(\phi_{3}^{1})_{1}}\right\rangle&=&
\left\langle\frac{\partial^{3}V}{\partial f_a\partial(\phi_{1}^{3})_{1}\partial(\phi_{3}^{1})_{2}}\right\rangle=
\left\langle\frac{\partial^{3}V}{\partial f_a\partial(\phi_{1}^{3})_{2}\partial(\phi_{3}^{1})_{1}}\right\rangle=
2\sqrt{2} e_3^a,\\ \nonumber\\
\left\langle\frac{\partial^{3}V}{\partial f_a\partial(\phi_{1}^{2})_{1}\partial(\phi_{2}^{1})_{1}}\right\rangle&=&
4 \sqrt{2} c_4^a\alpha_1,
\end{eqnarray}

\begin{eqnarray}
\left\langle\frac{\partial^{3}V}{\partial f_b\partial(\phi_{1}^{2})_{1}\partial(\phi_{2}^{1})_{2}}\right\rangle&=&
\left\langle\frac{\partial^{3}V}{\partial f_b\partial(\phi_{1}^{2})_{2}\partial(\phi_{2}^{1})_{1}}\right\rangle=
\left\langle\frac{\partial^{3}V}{\partial f_d\partial(\phi_{1}^{2})_{1}\partial(\phi_{2}^{1})_{1}}\right\rangle=
4 e_3^a,\\ \nonumber\\
\left\langle\frac{\partial^{3}V}{\partial (S_{1}^{2})_{1}\partial(\phi_{2}^{3})_{1}\partial(\phi_{3}^{1})_{1}}\right\rangle&=&
4 c_4^a(2\alpha_1-\alpha_3),\\ \nonumber\\
\left\langle\frac{\partial^{3}V}{\partial (S_{1}^{2})_{1}\partial(\phi_{2}^{3})_{1}\partial(\phi_{3}^{1})_{2}}\right\rangle&=&
\left\langle\frac{\partial^{3}V}{\partial (S_{1}^{2})_{1}\partial(\phi_{2}^{3})_{2}\partial(\phi_{3}^{1})_{1}}\right\rangle=
\left\langle\frac{\partial^{3}V}{\partial (S_{1}^{2})_{2}\partial(\phi_{2}^{3})_{1}\partial(\phi_{3}^{1})_{1}}\right\rangle=
-4 e_3^a.
\end{eqnarray}

\section{Decoupling limit of GLSM}\label{deli}

The four-quark fields are decoupled in the limit $d_2, e_{3}^{a}\rightarrow 0$ and $\gamma_1 \rightarrow 1$. To find five unknown parameters in this limit $(c_2, c_4^a,A_1,A_3, \alpha_1,\alpha_3)$, we use $m_\pi$, $m_a$ and $A_3/A_1$ as experimental inputs
\begin{eqnarray}
m_{\pi}^2&=&-2 c_2 +4 c_4^a \alpha_1^2,\nonumber \\
m_{a}^2&=&-2 c_2 +12 c_4^a \alpha_1^2,\nonumber \\
\frac{A_3}{A_1}&=& 20 \rightarrow 30,\nonumber \\
F_{\pi}&=& 2 \alpha _{1},
\end{eqnarray}
besides two minimum equations
\begin{eqnarray}
-2A_1+4c_a^a \alpha_1^3-2 \alpha_1 c_2&=&0\nonumber, \\
-2A_3+4c_a^a \alpha_3^3-2 \alpha_3 c_2&=&0.
\end{eqnarray}
The experimental input for the determination of  the remaining parameter $c_3$ is ${\rm Tr(M_\eta^2)}$
\begin{equation}
m_\eta^2+m_\eta^{'2}=-4 c_2-\frac{16 c_3 }{\alpha_1^2}+4 c_4^a \alpha_1^2-\frac{8 c_3 }{\alpha_3^2}+4 c_4^a \alpha_3^2.
\end{equation}
In this limit, the physical vertices are
\begin{eqnarray}\label{gf2pipi}
\gamma_{\sigma \pi \pi }&=&4 c_4^a \alpha_1,\nonumber \\
\gamma_{f_0 \pi \pi }&=&0,\nonumber \\
\gamma_{f_2 KK}&=& -4 \sqrt{2} c_4^a \alpha_1+8 \sqrt{2} c_4^a \alpha_3,\nonumber\\
\gamma_{\kappa \pi K}&=& 4 c_4^a \alpha^3, \nonumber\\
\gamma_{a_0 \pi \eta }&=&\frac{8\sqrt{2}\,c_3 \cos(\theta_p)}{\alpha_{1}^3}+4\sqrt{2}\,c_4^a\cos(\theta_p)\alpha_1-\frac{8\,c_3\sin(\theta_p)}{\alpha_{1}^2\alpha_3},\nonumber \\
\gamma_{a_0KK}&=&8 c_4^a \alpha_1-4 c_4^a \alpha_3,
\end{eqnarray}
where $\theta_p$ is the pseudoscalar mixing angle.

\begin{equation}
\left[
\begin{array}{cc}
\eta\\
\eta'
\end{array}
\right]
=
\left[
\begin{array}{cc}
\cos \theta_p & -\sin \theta_p\\
\sin \theta_p & \cos \theta_p
\end{array}
\right]
\left[
\begin{array}{cc}
\eta_a\\
\eta_b
\end{array}
\right].
\end{equation}

\end{document}